\documentclass[aps,prl,amsmath,twocolumn,groupedaddress,letterpaper,floatfix]{revtex4-1}
\usepackage{graphicx,color}
\usepackage{verbatim}
\usepackage{amssymb}   
\usepackage{amsmath}
\usepackage{amsfonts}
\usepackage{mathdots}
\usepackage{hyperref}
\usepackage{epsfig}

\usepackage{braket}
\usepackage{bm}
\begin{document}
	\title{Anomalous $h/2e$-periodicity and Majorana zero modes in chiral Josephson junctions}

        \author{Zi-Ting Sun}\thanks{These authors contributed equally to this work.}
        \author{Jin-Xin Hu}\thanks{These authors contributed equally to this work.}
         \author{Ying-Ming Xie}\thanks{yxieai@connect.ust.hk}
           \author{K. T. Law}\thanks{phlaw@ust.hk}
	\affiliation{Department of Physics, Hong Kong University of Science and Technology, Clear Water Bay, Hong Kong, China} 	
	
	\date{\today}
	\begin{abstract}
Recent experiments reported that quantum Hall chiral edge state-mediated Josephson junctions (chiral Josephson junctions) could exhibit Fraunhofer oscillations with a periodicity of either $h/e$ [Vignaud \textit{et al}.,~Nature~(2023)] or $h/2e$ [Amet \textit{et al}.,~Science~\textbf{352}~966~(2016)]. While the $h/e$-periodic component of the supercurrent had been anticipated theoretically before, the emergence of the $h/2e$-periodicity is still not fully understood. In this work, we show that the chiral edge states coupled to the superconductors become chiral Andreev edge states. In short junctions, the coupling of the chiral Andreev edge states can cause the $h/2e$-magnetic flux periodicity. Our theory resolves the long-standing puzzle concerning the appearance of the $h/2e$-periodicity in chiral Josephson junctions.  Furthermore, we explain that when the chiral Andreev edge state couple, a pair of localized Majorana modes appear at the ends of the Josephson junction, which are robust and independent of the phase difference between the two superconductors. As the $h/2e$-periodicity and the Majorana zero modes have the same physical origin, the Fraunhofer oscillation period can be used to identify the regime with Majorana zero modes.
	\end{abstract}
	\maketitle
 
\emph{Introduction.}---The topological boundary state-superconductor hybrids are promising platforms for realizing Majorana zero modes and topological quantum computation \cite{fu2008superconducting,nayak2008non,qi2010chiral,alicea2012new,mourik2012signatures,beenakker2013search,chen2018quasi,yazdani2023hunting}. In the past decade,  Josephson junctions with supercurrents mediated by edge states had been achieved in various topological systems, such as in quantum Hall insulators \cite{ma1993josephson,calado2015ballistic,amet2016supercurrent,draelos2018investigation,guiducci2019toward,seredinski2019quantum,vignaud2023evidence}, the quantum spin Hall insulator HgTe/HgCdTe quantum wells \cite{hart2014induced} and the higher order topological insulators \cite{schindler2018higher,choi2020evidence,bernard2023long}. A prominent feature of edge state-mediated Josephson junctions is to give rise to superconducting quantum interference (SQUID) like oscillations in the Fraunhofer patterns \cite{hart2014induced,baxevanis2015even,de2018h,vigliotti2022anomalous} as illustrated in the right panels of Fig.~\ref{fig:fig1}.

\begin{figure}
		\centering
		\includegraphics[width=1\linewidth]{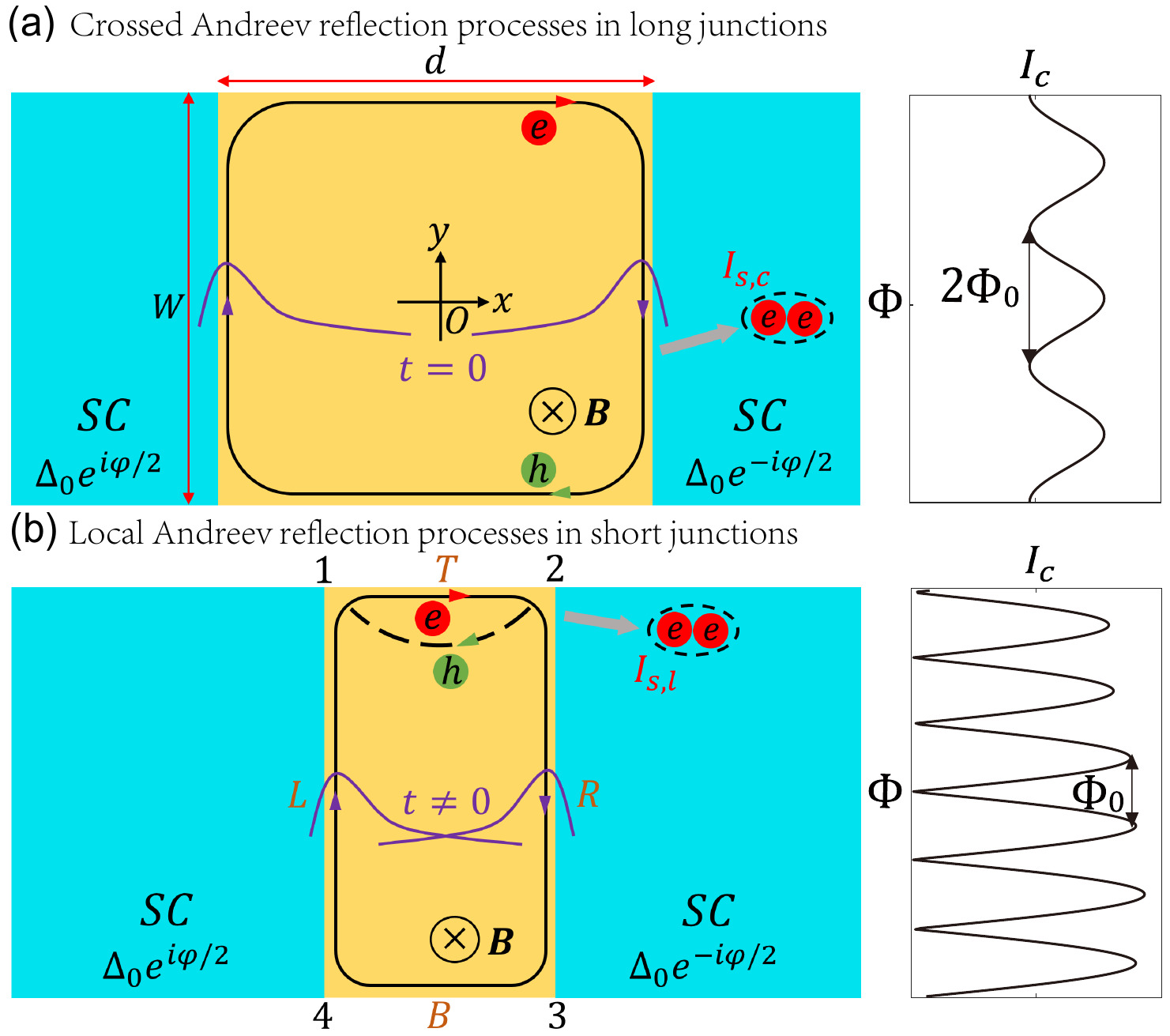}
		\caption{Origins of 2$\Phi_0$- and $\Phi_0$-periodicity in chiral Josephson junctions. (a) In the long junction, the crossed Andreev reflection processes dominate. Namely, an electron \textit{e} from the top edge (T) is reflected as a hole \textit{h} in the bottom edge (B). This results in a Cooper pair \textit{e-e} entering the superconductor (SC). The critical Josephson current $I_c$ as a function of magnetic flux $\Phi$ has a period of $2\Phi_0$ (right panel).
  (b) In the short junction, the wavefunctions (illustrated as purple lines) of the left edge state (L) and the right edge state (R) have a finite coupling $t$. In this case, the supercurrent is mainly mediated by the local Andreev reflections, namely, an electron \textit{e} from the top edge is reflected locally as a hole \textit{h}, which results in a Fraunhofer pattern with a period of $\Phi_0$ instead (right panel).}
		\label{fig:fig1}
\end{figure}

Josephson junctions with chiral edge states (CESs) as the weak links are particularly interesting \cite{van2011spin,nakai2021edge,tang2022vortex,kurilovich2023disorder}. In these chiral Josephson junctions, the supercurrents are medicated by chiral edge states. It was anticipated by theories that chiral Josephson junctions exhibit $2\Phi_0=h/e$-periodicity in the Fraunhofer patterns, where $\Phi_0=h/2e$ is the magnetic flux quantum \cite{van2011spin,stone2011josephson,liu2017superconductor,alavirad2018chiral,dominguez2022fraunhofer,kurilovich2023disorder}. As depicted in Fig.~\ref{fig:fig1}~(a), due to the chiral nature of the CESs, in long junctions, the Andreev reflections can happen only when an electron from the top (T) edge hits the CES/superconductor interface is reflected as a hole along the bottom (B) edge [T and B label the edges as in Fig.~\ref{fig:fig1}~(b)]. Such Andreev reflections involving two CESs in a single tunnelling process are crossed Andreev reflections. The $2\Phi_0$-period of the Fraunhofer oscillations is twice the period of conventional SQUIDs involving Cooper pair interferences and is experimentally observed very recently [Vignaud \textit{et al}.~\cite{vignaud2023evidence}]. However, in another experiment by [Amet \textit{et al}.~\cite{amet2016supercurrent}], $\Phi_0$-periodicity was observed. Previous theoretical works reported in Ref.~\cite{liu2017superconductor} and Ref.~\cite{alavirad2018chiral} have noticed such inconsistency between theories and the experiments, but it remains a puzzle how the $\Phi_0$-periodicity emerges in a chiral Josephson junction~\cite{amet2016supercurrent}. Therefore, a theory reconciling the aforementioned discrepancies between experiments and theories is highly desirable at the current stage.

In this work, we revisit the Fraunhofer patterns in chiral Josephson junctions. Unlike previous theories, we take into account the possible hybridization between the chiral edge states (or more precisely, the chiral Andreev edge states introduced later) along the two quantum (or anomalous) Hall/superconductor interfaces (see the left panels of Fig.~\ref{fig:fig1}). Using an edge-channel model as well as a lattice model, we unambiguously show that such hybridization would result in a crossover of the period in Fraunhofer oscillations from 2$\Phi_0$ to $\Phi_0$ as the junction length decreases. Specifically, we find that the supercurrents originated from crossed Andreev reflections between two separated edges [see Fig.~\ref{fig:fig1}~(a)] give rise to 2$\Phi_0$ oscillations, while the supercurrents originated from local Andreev reflections [see Fig.~\ref{fig:fig1}~(b)] exhibit $\Phi_0$ oscillations. Importantly, we demonstrate that in the $\Phi_0$-periodicity regime, robust Majorana zero-energy modes (MZMs) appear at the Josephson junction without fine-tunings. Our theory suggests that short superconductor/quantum anomalous Hall insulator (QAHI)/superconductor Josephson junctions are promising platforms for realizing MZMs (Fig.~\ref{fig:fig3}).

\emph{Edge-channel model.}---We first construct an edge-channel model to capture the low-energy physics of Josephson junctions with CESs in the weak link as depicted in Fig.~\ref{fig:fig1}. This model is applicable to describe Josephson junctions with a quantum Hall or a QAHI weak link. In this model, the two superconductors have a bulk pairing potential $\Delta_0$ and a phase difference $\varphi$. The electrons of the CESs propagate along the clockwise direction. Thus the model Hamiltonian can be written as:
\begin{equation}
\label{Eq1}
H=\sum_{\gamma}\int d\bm{r} \Psi_{\gamma}^{\dagger}(\bm{r}) H^{B d G}_\gamma(\bm{r}) \Psi_{\gamma}(\bm{r}),
\end{equation}
where $\gamma$ labels the left/right (L/R) and top/bottom (T/B) edges. The Nambu basis vector $\Psi_\gamma(\bm{r})=\left[ \psi_{\uparrow \gamma}(\bm{r}), \psi^{\dagger}_{\downarrow \gamma}(\bm{r})\right]^T$, where $\psi_{s \gamma}(\bm{r})$ is an annihilation operator for an electron with spin $s$ at $\gamma$ edge at position $\bm{r}$. The Bogoliubov-de-Gennes (BdG) Hamiltonian reads
\begin{equation}
\label{Eq2}
H^{BdG}_{\gamma}(\bm{r})=H^{B d G}_{0,\gamma}(\bm{r})+H^{B d G}_{1}(\bm{r}).
\end{equation}
Here, $H^{B d G}_{0,\gamma}(\bm{r})$ is the propagating Hamiltonian of four edges in Fig.~\ref{fig:fig1}~(a), while $H^{B d G}_{1}(\bm{r})$ characterize the coupling between the left and the right CESs [Fig.~\ref{fig:fig1}~(b)]. The exact form of $H^{B d G}_{1}(\bm{r})$ is not a concern here. As it is shown later, $H^{B d G}_{1}(\bm{r})$ is responsible for giving rise to the $h/2e$-periodicity in the Fraunhofer oscillations. In the presence of an out-of-plane magnetic field, the propagating Hamiltonian yields
\begin{equation}
\label{Eq3}
H^{B d G}_{0,\gamma}=\left(\begin{array}{cc}
  \bm{v_{\gamma} \cdot (\hat{\bm{p}}}+e\bm{A})-\mu_{\gamma} & \Delta_{\gamma} e^{i \sigma_\gamma\frac{\varphi }{2} } \\
\Delta_\gamma e^{-i \sigma_\gamma\frac{\varphi }{2} } & \bm{v_{\gamma} \cdot (\hat{\bm{p}}}-e\bm{A})+\mu_{\gamma}
\end{array}\right).
\end{equation}
Here, the momentum operator is $\hat{\bm{p}}=- i\hbar\nabla$, $\Delta_{\gamma}$ is the effective pairing potential ($\Delta_{\gamma=\text{T}/\text{B}}=0$, $\Delta_{\gamma=\text{L}/\text{R}}=\Delta$), $\mu_{\gamma}$ is the chemical potential ($\mu_{\gamma=\text{L}/\text{R}}=\mu',\mu_{\gamma=\text{T}/\text{B}}=\mu$), $\bm{v_{\gamma}}$ denotes the Fermi velocity of edge states ($\bm{v_{\gamma=\text{T/B}}}=(\sigma_{\gamma} v_\text{F}, 0)$, $\bm{v_{\gamma=\text{L/R}}}=(0, \sigma_{\gamma} v_\text{s})$ with $\sigma_{\text{L/R}}=\sigma_{\text{T/B}}=\pm 1$), $v_F$ is the bare velocity of CESs, and $v_s$ is the renormalized Fermi velocity. Because of the proximity effects, in which the electrons enter the superconductors virtually, $v_\text{s}$ would be approximately renormalized as $v_\text{s}\approx v_\text{F}/(1+g/\Delta_0)$ and the proximity pairing potential $\Delta\approx g\Delta_0/(g+\Delta_0)$ with $g$ as an effective coupling strength \cite{alavirad2018chiral,vignaud2023evidence}. The Landau gauge $\bm{A}=-yB\hat{\bm{x}}$ captures the effects of a magnetic field with strength $B$.

\emph{Possible Andreev reflection processes in short and long junctions.---} Before evaluating the supercurrent mediated by the CESs in Fig.~\ref{fig:fig1}, we first analyze the possible Andreev reflection processes. Such processes are significantly different in short and long junction limits. In the long junction limit where the coupling Hamiltonian $H_{1}^{BdG}(\bm{r})$ is negligible, the electrons and holes on the CESs propagate circularly along the edges. It is worth noting that the electrons and holes propagate in the same direction on a CES. As a result, for an Andreev process, an incoming electron at the top edge can only be reflected as a hole at the bottom edge [see Fig.~\ref{fig:fig1}~(a)]. Such crossed Andreev reflections result in a Cooper pair tunnelling from the top edge to the superconductor on the right, leading to a supercurrent denoted by $I_{s,c}$. On the other hand, in the short junction case, an additional Andreev reflection path could appear. As schematically shown in Fig.~\ref{fig:fig1}~(b), an incoming electron at the top edge can be directly reflected as a hole to the left edge because of the finite coupling between the left and right CESs. Such local Andreev reflections at one edge give rise to a supercurrent denoted by $I_{s,l}$. In the following section, by using the scattering matrix method \cite{beenakker1991universal}, we demonstrate that $I_{s,l}$ and $I_{s,c}$ have periods $\Phi_0$ and $2\Phi_0$ respectively. Our theory provides a unified description for both experiments reported in \cite{amet2016supercurrent} and \cite{vignaud2023evidence}.

\emph{Scattering matrix for chiral Josephson junctions.---}To derive the scattering matrix, we first analyze the scattering modes given by the eigenstates of $H_{0,\gamma}^{BdG}$. Specifically, the scattering modes behave as plane waves of pure electrons and holes on the top and bottom edges. On the other hand, on the left and right edges, the scattering modes are superpositions of electrons and holes, which can be written as
\begin{equation}
\label{Eq4}
\Psi^S_{e \gamma}=\zeta_{e\gamma} e^{i  k_{s,e,\gamma} y}, \,
\Psi^S_{ h\gamma}=\zeta_{h\gamma} e^{i  k_{s, h,\gamma} y}
\end{equation}
with the wavevecor $k_{s, e/h, \gamma}=\left(\epsilon\pm\sqrt{\Delta^2+\mu^{\prime 2}}\right)/\sigma_\gamma \hbar v_s$, $\epsilon$ is the energy, $e \, (h)$ represents the electron (hole)-like state. $\zeta_{e\gamma}=(e^{i \sigma_\gamma \varphi/2}, -e^{-\beta})_\gamma^T$, $\zeta_{h\gamma}=(e^{-\beta}, e^{-i \sigma_\gamma \varphi/2})_\gamma^T$, with $\beta=\operatorname{arcsinh} (\mu^{\prime}/\Delta )$. The subscript $\gamma$ and $\sigma_\gamma$ are defined in the same way as in the BdG Hamiltonian. These chiral modes, which are superpositions of electrons and holes, were called chiral Andreev edge states (CAES) in previous experiments \cite{lee2017inducing,zhao2020interference}. As will be shown later, the CAES are essential for forming MZMs in the chiral Josephson junctions.

However, the CAESs become non-chiral due to the hybridization effect of the wavefunctions when the junction length $d$ is comparable with the localization length $\xi_d$ of the CAES, i.e., the coupling Hamiltonian $H_{1}^{BdG}$ becomes essential. In this case, the left and right edges are coupled, and we can rewrite the two scattering modes 
\begin{equation}
\label{Eq5}
\begin{split}
    \Psi^S_{\alpha +}&=[ r_{\alpha,LL}\zeta_{\alpha L}+t_{\alpha,LR}\zeta_{\alpha  R}]  e^{i  k_{s,\alpha,+} y},\\
    \Psi^S_{\alpha -}&=[ r_{\alpha,RR}\zeta_{\alpha R}+t_{\alpha,RL}\zeta_{\alpha L}]  e^{i  k_{s,\alpha,-} y},
\end{split}
\end{equation}
where $\pm$ labels the propagating direction, $\alpha=e/h$, and $\sigma_{e/h}=\pm 1$. We assume the dispersion relation of the scattering modes does not change as an approximation, i.e., $k_{s, \alpha, \pm}=\pm\left(\epsilon+\sigma_\alpha\sqrt{\Delta^2+\mu^{\prime 2}}\right)/ \hbar v_s$. The overlap of the left and right edges is characterized by $t_{\alpha, LR(RL)}=e^{-\frac{d}{\xi_d}\mp i \sigma_\alpha \pi \frac{\Phi}{\Phi_0}\frac{y}{W}}$, which is dressed by Peierls substitution and is exponentially suppressed when $d/\xi_d$ increases. Here, the magnetic flux $\Phi=BdW$ with $W$ as the junction width. Note that the unitary condition requires $|r_{\alpha,\gamma\gamma}|^2+|t_{\alpha,\gamma\gamma'}|^2=1$ and $r_{\alpha,\gamma\gamma}=\sqrt{1-\text{exp}(-2d/\xi_d)}$ is assumed.

 The scattering processes can be manifested by investigating how incoming and outgoing states at four corners are related. By properly formulating, we can obtain $\psi_{\text {in}}=\mathcal{S}_{N} \psi_{\text {out}}$, $\psi_{\text {out}}=\mathcal{S}_{A} \psi_{\text {in}}$, where the scattering states are defined as $\psi_{\text {in }}=[c_{e}^{\prime}\left(4\right), c_{h}^{\prime}\left(4\right), c_{e}\left(2\right), c_{h}\left(2\right)]^T$,
$\psi_{\text {out }}$=$[c_{e}\left(1\right), c_{h}\left(1\right),c_{e}^{\prime}\left(3\right),c_{h}^{\prime}\left(3\right)]^T$ (1 to 4 label the corner positions at Fig.~\ref{fig:fig1}~(b)). The scattering matrix $S_N$ describes the normal transition from 1 to 2 (3 to 4) at the top (bottom) edge, while $S_A$ describes how an incoming scattering state at 2 (4) becomes an outgoing state at 3 (1) through side edges. The explicit forms of $S_{N}$ and $S_{A}$ are shown in Note~1~B of Supplemental Material \cite{NoteX}.

\begin{figure}
		\centering
		\includegraphics[width=1.0\linewidth]{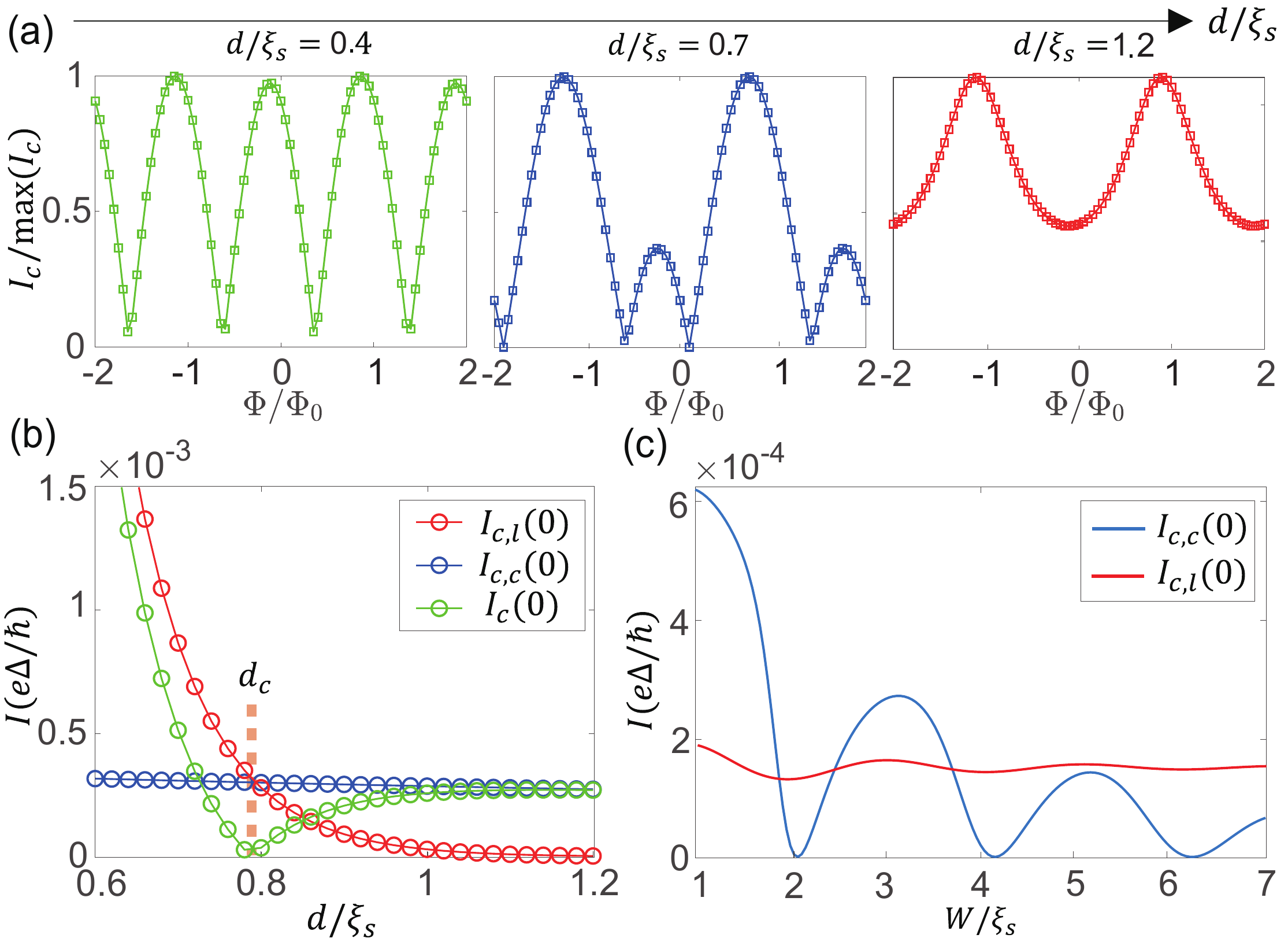}
		\caption{(a) The crossover from 2$\Phi_0$ oscillations to $\Phi_0$ oscillations with the increase of junction length $d$. (b) The total critical supercurrent $I_{c}$, the critical supercurrent from local Andreev reflections $I_{c,c}$, and the critical supercurrent from local Andreev reflections $I_{c,l}$ at zero magnetic flux as a function of $d$. $d_c$ labels the critical length where $I_{c,c}\approx 0$. In (a) and (b), the width of the junction is set as $W/\xi_s=3$. (c) Junction width dependence of $I_{c,c}$ and $I_{c,l}$ for $d/\xi_s=0.84$. The width regions are highlighted, where $I_{c,c}$ is larger than $I_{c,l}$. In these calculations, we set the parameters as $v_F=2$, $\Delta_0=0.08$, $\mu=0.02$, and $\mu'=0.1$. The coupling strength $g=\Delta_0/5$, which yields $v_s=1.6$ and $\Delta=\Delta_0/6$. The temperature is $k_B T=0.05\Delta_0$. To describe the coupling of the edge states, the parameter $\xi_d$ is set to be 0.1$\xi_s$, where $\xi_s$ is the superconductor coherent length. }
		\label{fig:fig2}
\end{figure}

\emph{Origin of $\frac{h}{e}$- and $\frac{h}{2e}$-periodicity.---}The flux-dependent Josephson current through the junction is related to the scattering matrix by \cite{brouwer1997anomalous}
\begin{equation}
\label{Eq6}
  I_s(\varphi,\Phi)=-\frac{2e k_B T}{\hbar} \frac{d}{d \varphi} \sum_{n=0}^{\infty} \ln \operatorname{det}\left[1-\mathcal{S}_A(i \omega_n) \mathcal{S}_N(i \omega_n)\right], 
  \end{equation} 
  where $\omega_n=(2 n+1)\pi k_B T $ are fermionic Matsubara frequencies, $T$ is the temperature. The total supercurrent $I_s(\varphi,\Phi)\approx I_{s,c}(\varphi,\Phi)+I_{s,l}(\varphi,\Phi)$, where we define $I_{s,c}$ and $I_{s,l}$ as the supercurrent arising from the crossed and the local Andreev reflections respectively are shown in Fig.~\ref{fig:fig1}.

We now study the critical supercurrents when $\varphi$ varies from 0 to 2$\pi$ at a fixed $\Phi$, which gives rise to the Fraunhofer pattern. Before looking at the total critical supercurrents $I_c(\Phi)=\text{max}_{\varphi}\{I_s(\varphi,\Phi)\}$, we first study the critical currents given by the crossed and the local Andreev reflections respectively. The critical supercurrents $I_{c,c}=\text{max}_{\varphi}\{I_{s,c}(\varphi,\Phi)\}$, and $I_{c,l}=\text{max}_{\varphi}\{I_{s,l}(\varphi,\Phi)\}$ are (details derivations can be found in Supplemental Material Note~1~C \cite{NoteX}): 
\begin{eqnarray}
\label{Eq7}
I_{c,c}(\Phi)=|I_0+I_1\cos(\pi\frac{\Phi}{\Phi_0}+\phi)|, \\
I_{c,l}(\Phi)=|I_2\cos(\pi\frac{\Phi}{\Phi_0}+\phi')| \label{Eq8},
\end{eqnarray}
where $I_0=(2ek_BT/\hbar)\sin^2(\delta k W)\text{sech}\Gamma$, $I_1=(e k_BT/2\hbar)\sin^2(2\delta k W)\text{sech}^2\Gamma$ with $\Gamma=2\pi k_BT(d/\hbar v_F+W/\hbar v_s)$ as the ratio between temperature and Thouless energy, $\delta k=\sqrt{\Delta^2+\mu'^2}/ \hbar v_s$ as the Fermi wavevector, $I_{2}=(4 e k_BT/\hbar) e^{- (\frac{\pi k_BT d}{\hbar v_\text{F}}+\frac{d}{\xi_d})}$, and the field-independent phase $\phi=2\mu d/ \hbar v_F+\text{arctan}(2\beta \tan\delta k W), \phi'=\mu d/\hbar v_F$. 

First of all, one can show that $I_0$ is always larger than $I_1$ in $I_{c,c}(\Phi)$, thus the periodicity of $I_{c,c}(\Phi)$ is $2\Phi_0$. This is consistent with previous theoretical findings \cite{van2011spin,alavirad2018chiral}. Importantly, the critical current $I_{c,l}(\Phi)$ possesses a period of $\Phi_0$ Fraunhofer oscillations. Physically, the hopping between the two edge states enables the supercurrent to enter the superconductors via the top and bottom edge supercurrent independently, and the interference between the supercurrents at two edge states thus acquires a period of $\Phi_0$, which mimics the scenario of the quantum spin Hall Josephson junction \cite{hart2014induced}. In other words, the local Andreev reflection induced by the left-right edge coupling originates the $\Phi_0$ periodicity of the Fraunhofer pattern.

To see the crossover from $\Phi_0$-periodicity to $2\Phi_0$-periodicity as the junction length increases, we numerically evaluate Eq.~\ref{Eq6} without approximations. The calculated Fraunhofer patterns at various lengths with $d/\xi_s=0.4, 0.7, 1.2$ ($\xi_s=\hbar v_F/\Delta_0$ as superconducting coherence length) are shown in Fig.~\ref{fig:fig2}~(a). As expected, in the long junction region when $d/\xi_s = 1.2$, the coupling between the left and right CESs is weak and the crossed Andreev reflection processes dominate. As a result, the $2\Phi_0$-periodic component is dominant. On the other hand, in the short junction limit when $d/\xi_s=0.4$, the local Andreev reflection contribution becomes more significant and the $\Phi_0$-periodic component is dominant. 

 Fig.~\ref{fig:fig2}~(b) shows the critical currents in the zero magnetic field limit ($\Phi=0$) as a function of junction length $d$, where $I_{c,c}$ ($I_{c,l}$) only includes crossed (local) Andreev reflection contributions and $I_{c}$ is the total critical supercurrent. As expected, $I_{c,l}$ decreases more dramatically with the increase of $d$ compared with $I_{c,c}$. Notably, the $I_{c,c}$ and $I_{c,l}$ crossover each other at a critical length $d_c$. When $d<d_c$, the $\Phi_0$-periodic component can dominate. Fig.~\ref{fig:fig2}~(c) shows the width dependence of the critical current.  By fixing $d=0.84\xi_s$ and increasing $W$, it can be seen that $I_{c,l}$ is not sensitive to the change of $W$ but $I_{c,c}$ is exponentially suppressed as the width of the junction increases. This is consistent with the recent experiment that the $2\Phi_0$-periodicity was only observed in junctions with small $W$ \cite{vignaud2023evidence}. Finally, to further support our theoretical analysis, we adopt a minimal two-band QAHI lattice model in a superconductor/CI/superconductor configuration to simulate the Fraunhofer pattern with the lattice Green's function method \cite{furusaki1994dc,ando1991quantum,asano2001numerical,asano2003josephson}. The lattice model results are fully consistent with the results shown in Fig.~\ref{fig:fig2} of the edge-channel model (for details of the lattice model, see Supplemental Material Note~2~G \cite{NoteX}). Remarkably, taking the parameters from the two experiments \cite{amet2016supercurrent,vignaud2023evidence}, we show that the period of oscillations in both experiments can be explained well by our theory (details can be found in Supplemental Material Note~1~E \cite{NoteX}).

\begin{figure}
		\centering
		\includegraphics[width=1.0\linewidth]{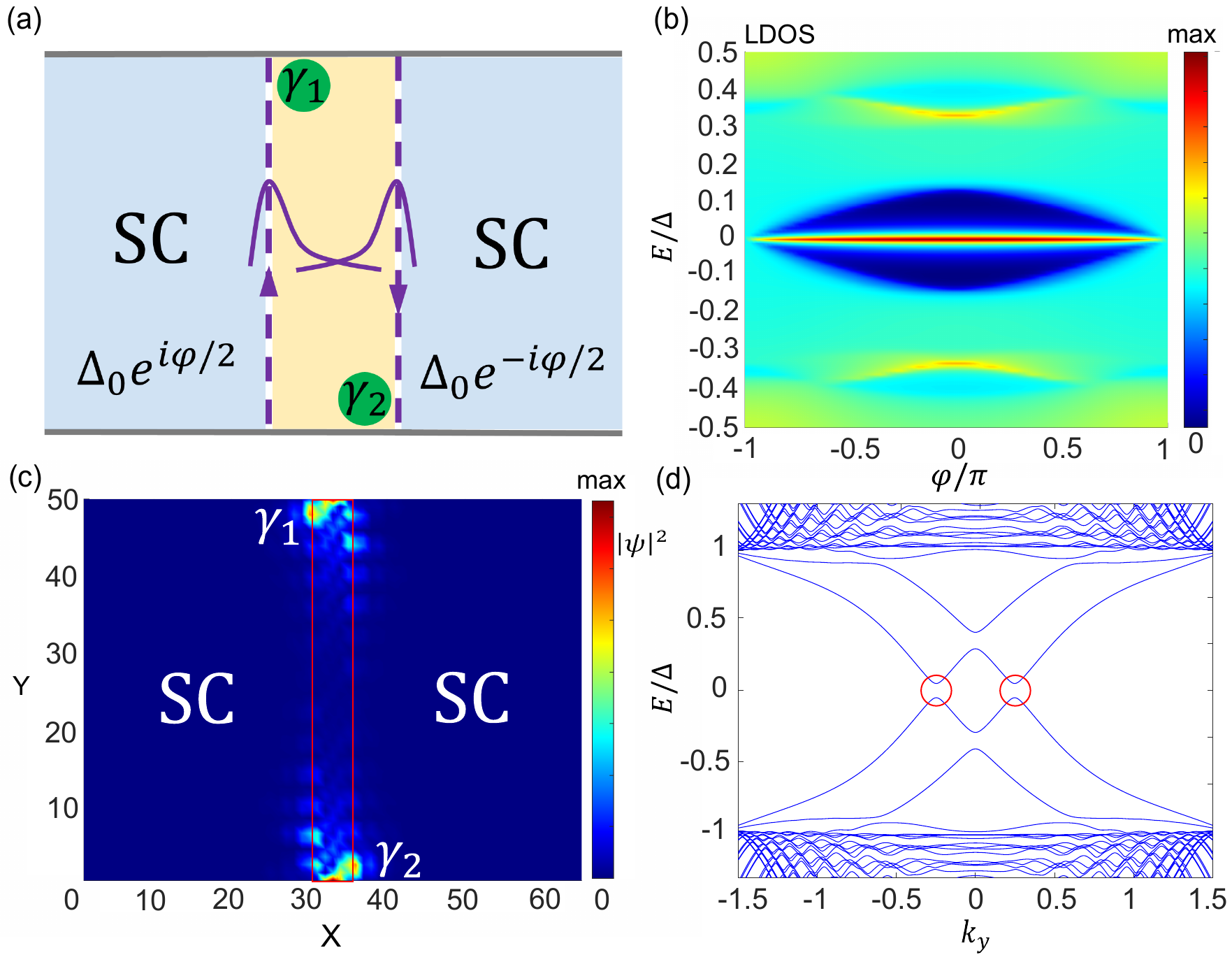}
		\caption{MZMs in the superconductor-QAHI-superconductor junction simulated using a lattice model: (a) The schematic picture of a short Josephson junction with counter-propagating left and right edge states (labelled as purple dashed lines). When the left and right edge state wavefunctions overlap (depicted as solid purple lines), MZMs emerge at the ends of the weak link (labelled as $\gamma_1$ and $\gamma_2$). (b) The LDOS at one end of the junction versus the phase difference $\varphi$. There is clearly a zero energy mode at the junction, which is independent of $\varphi$. (c) The localized wavefunction of MZMs at the Josephson junction (the boundary of the junction region is highlighted by a red box). (d) The energy spectrum of superconductor-QAHI-superconductor junction, with the periodic boundary condition along the $y$-direction. The four CAESs emerge in the low-energy states. The energy gaps circled in red are the effective pairing gaps of the states at the Josephson junction. (c) and (d) correspond to $\varphi=0$ in (b). See the details of the lattice model and other parameters in Supplementary Note 2.~G \cite{NoteX}.}
		\label{fig:fig3}
\end{figure}

\emph{Emergence of MZMs in short junctions.}---In the above sections, it was shown that the anomalous periodicity of $h/2e$ is induced by the left-right edge coupling. In this section, we demonstrate another important consequence of the left-right edge coupling, namely, the emergence of robust MZMs at the Josephson junction without fine-tuning. The schematic picture of a short Josephson junction with counter-propagating left and right edge modes is depicted in Fig.~\ref{fig:fig3}~(a). A superconductor/QAHI/superconductor junction lattice model is built in the Supplemental Material Note~2~G \cite{NoteX}. If the left and right edge modes are coupled, we can regard the QAHI weak link as a quasi-1D wire with a single helical channel \cite{chen2018quasi,xie2020creating,law2022creating}. When this helical channel is coupled to a superconductor, MZMs emerge at the two ends of the quasi-1D channel \cite{kitaev2001unpaired,oreg2010helical,sau2010generic,lutchyn2010majorana,lutchyn2011search,choy2011majorana,klinovaja2013topological}.   

To demonstrate the existence of MZMs in this system, the local density of states (LDOS) at a transverse end of the junction as a function of the Josephson phase difference $\varphi$ is plotted in Fig.~\ref{fig:fig3}~(b). The prominent LDOS peak near the zero energy indicates the MZM, which resides inside the pairing gap and is robust against the variation of the phase difference $\varphi$  (Fig.~\ref{fig:fig3}~(b)). Notably, the optimal phase difference is $\varphi=0$ to maximize the topological superconducting gap. This is in sharp contrast with the previous topological Josephson junction proposals in which the topological regime appears when the phase difference of the junction is near $\pi$ \cite{pientka2017topological,hell2017two,ke2019ballistic,fornieri2019evidence,ren2019topological,xie2023gate}. It is also important to note that the topological superconducting gap protecting the MZM is quite sizable, which is about 10\% of the superconducting gap of the superconductors (and it can be further optimized by increasing the coupling between the CAESs from the left and right edges). The wavefunction of the MZMs is shown in Fig.~\ref{fig:fig3}~(c), which displays the expected localization behaviour at the two ends of the junction. 

Fig.~\ref{fig:fig3}~(d) shows the energy spectrum of the Josephson junction. The propagating states inside the pairing gap are the CAESs. When the CAESs from the left and the right edges couple, hybridization gaps emerge, which are highlighted by the red circles. When the hybridization gap is finite, a pair of MZMs emerge at the two ends of the Josephson junction with wavefunctions depicted in Fig.~\ref{fig:fig3}~(c). This is the second main result of this work. Interestingly, the existence of $h/2e$ Fraunhofer oscillations and the existence of the MZMs originate from the same mechanism in our theory, which is the coupling of the CAESs. Therefore, the $h/2e$ oscillations in the Fraunhofer pattern can indicate the presence of MZMs in chiral Josephson junctions (given that the number of chiral edge states at the junction is odd). 

\emph{Conclusion and discussion.}---In this work, we have discovered that an anomalous $\Phi_0$ periodic supercurrent is generated by the coupling of CAESs on opposite edges in CESs-mediated Josephson junctions (chiral Josephson junction). Importantly, the coupling of CAESs is also responsible for the appearance of MZMs at the Josephson junction, as long as the number of CESs at each edge of the weak link is odd. Therefore, by measuring the $\Phi_0$-periodicity of the Fraunhofer pattern, one can identify the parameter regime with MZMs.

Recently, the superconductor/QAHI/superconductor Josephson junction has been realized in the gate-defined Josephson junction in twisted bilayer graphene \cite{diez2023symmetry}, in which the superconducting states, as well as the QAHI state, can be achieved in a single piece of twisted bilayer graphene sample through electric gating \cite{sharpe2019emergent,serlin2020intrinsic,diez2023symmetry,xie2023varphi,hu2023josephson}. And the model in Fig.~\ref{fig:fig3}~(a) can also be achieved by depositing superconducting electrodes on moir\'{e} transition metal dichalcogenides, where robust QAHI states have been experimentally observed \cite{li2021quantum,cai2023signatures,zeng2023thermodynamic,park2023observation,xu2023observation}. These newly discovered QAHI platforms are promising for observing the 2$\Phi_0$ period to $\Phi_0$ period crossover and realizing Majorana zero modes. 

\emph{Acknowledgements}---We thank Carlo Beenakker for inspiring discussions. K. T. L. acknowledges the support of the Ministry of Science and Technology, China, and Hong Kong Research Grant Council through Grants No. 2020YFA0309600, No. RFS2021-6S03, No. C6025-19G, No. AoE/P-701/20, No. 16310520, No. 16310219, No. 16307622, and No. 16309718. Y.-M. X. acknowledges the support of Hong Kong Research Grant Council through Grant No. PDFS2223-6S01.

\bibliographystyle{apsrev4-1}
\bibliography{ref}

		\clearpage
		\onecolumngrid
\begin{center}
		\textbf{\large Supplemental Material for ``Anomalous $h/2e$-periodicity and Majorana zero modes in chiral Josephson junctions''}\\[.2cm]
		 Zi-Ting Sun,$^{1}$  Jin-Xin Hu,$^{1}$   Ying-Ming Xie,$^{1}$  K. T. Law$^{1}$\\[.1cm]
		{\itshape ${}^1$Department of Physics, Hong Kong University of Science and Technology, Clear Water Bay, Hong Kong, China}
\end{center}

\date{\today}
	\maketitle

\setcounter{equation}{0}
\setcounter{section}{0}
\setcounter{figure}{0}
\setcounter{table}{0}
\setcounter{page}{1}
\renewcommand{\theequation}{S\arabic{equation}}
\renewcommand{\thesection}{ \Roman{section}}

\renewcommand{\thefigure}{S\arabic{figure}}
 \renewcommand{\thetable}{\arabic{table}}
 \renewcommand{\tablename}{Supplementary Table}

\renewcommand{\bibnumfmt}[1]{[S#1]}
\renewcommand{\citenumfont}[1]{S#1}
\makeatletter
\maketitle

\section*{\bf{\uppercase\expandafter{NOTE 1: scattering matrix method}}}
\subsection{A. Scattering matrices in the long junction case}
\begin{figure}
  \centering
  \includegraphics[width=0.9\linewidth]{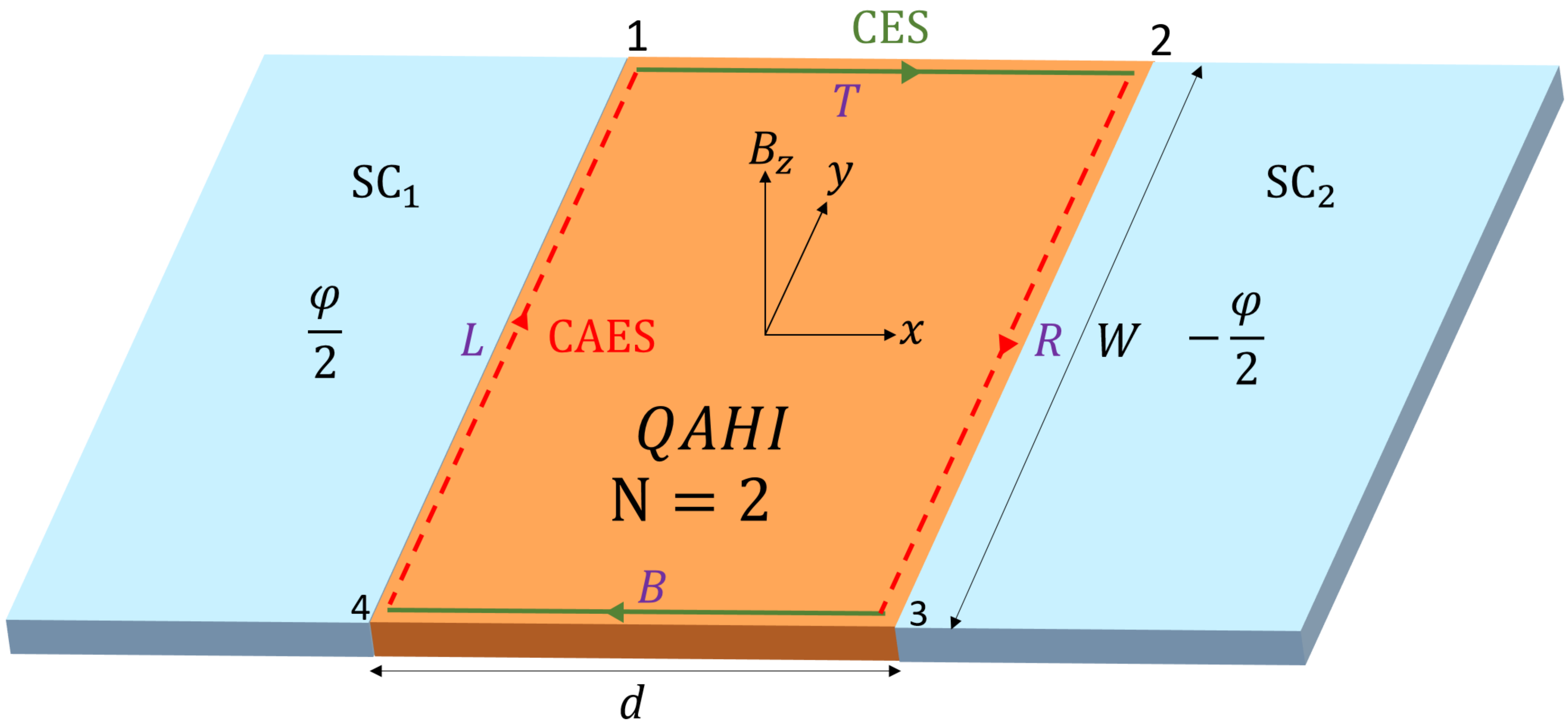}
  \caption{Configuration of the system, comprising an N=2 quantum anomalous Hall (QAH)
	weak link between two SC electrodes with a phase difference $\varphi$. The spin-degenerate chiral edge modes propagate along the clockwise direction on the boundary contour. Abbreviations: QAHI (quantum anomalous Hall insulator), CES (chiral edge state), CAES (chiral andreev edge state), SC (superconductor).}
  \label{fig:figS1}
\end{figure}

We start with a geometry depicted in Fig.~\ref{fig:figS1}. On the edges without pairing potential (i.e., the top (T) and bottom (B) edges with $y=W/2$ and $y=-W/2$, respectively), the continuum Hamiltonian
describing the spin degenerate CESs in an external out-of-plane magnetic field reads
\begin{equation}
  H=\sum_{\alpha=\text{T},\text{B}}\int d x \Psi_\alpha^{\dagger}(x)\left( -i\sigma_{\alpha} \hbar v_F\tau_0 \partial_x  + e \sigma_{\alpha}  v_F \bm{A} \cdot \hat{\bm{x}} \tau_z -\mu \tau_z \right) \Psi_\alpha(x), \quad x \in\left[-\frac{d}{2}, \frac{d}{2} \right]
  \end{equation}
with Numbu basis $\Psi_\alpha(x)=\left[\psi_{\alpha}(x), \psi_{\alpha}^{\dagger}(x)\right]^T$, flux quantum $\Phi_0=h/2e$, Landau gauge $\bm{A}=-yB\hat{\bm{x}}$, and Pauli matrices $\tau$ operate in particle-hole space.
The chiral scattering states are given by
\begin{equation}
\begin{aligned}
\Psi_{e\alpha} &=\left(\begin{array}{l}
1 \\
0
\end{array}\right) e^{i \sigma_{\alpha} \left(k_{F, e}+\frac{\pi B W }{2\Phi_0}\right) x} \\
\Psi_{h\alpha} &=\left(\begin{array}{l}
0 \\
1
\end{array}\right) e^{i \sigma_{\alpha} \left(k_{F, h}-\frac{\pi B W }{2\Phi_0}\right) x}
\end{aligned}
\end{equation}
with $\sigma_{T/B}=\pm 1$, and $k_{F, e/h} =(\epsilon\pm \mu)/\hbar v_F$,
where $v_F$ is the bare Fermi velocity, $\mu$ is the chemical potential, $\epsilon$ denotes the energy of the edge states.
The two states $\Psi_{e\alpha}$ and $\Psi_{h\alpha}$ represent the electron and hole at the same edge. It is worth noting that the electron-type and hole-type states have the same Fermi velocity, which implies they move in the same direction.

In the region proximate to the SC, the low-energy effective Hamiltonian for the edge states at the left and right edges ($x=\pm \frac{d}{2}$) is
\begin{equation}
H=\sum_{\gamma=L,R}\int d y \Psi_{\gamma}^{\dagger}(y) H^{B d G}_{\gamma}(y) \Psi_{\gamma}(y), \quad y \in\left[-\frac{W}{2}, \frac{W}{2} \right]
\end{equation}
with BdG Hamiltonian
\begin{equation}
H^{B d G}_\gamma(y)=\left(\begin{array}{cc}
-i\sigma_{\gamma} \hbar v_{s} \partial_y-\mu^{\prime} & \Delta e^{i \sigma_{\gamma}\frac{\varphi }{2} } \\
\Delta e^{-i \sigma_{\gamma}\frac{\varphi }{2} } & -i\sigma_{\gamma} \hbar v_{s}\partial_y+\mu^{\prime}
\end{array}\right),
\end{equation}
where $\sigma_{L/R}=\pm 1$, $\Psi_\gamma(y)=]\psi_\gamma(y), \psi^{\dagger}_\gamma(y)]^T$ is the Nambu basis, $\mu^{\prime}$ is the chemical potential for the SC. $v_\text{s}\approx v_\text{F}/(1+g/\Delta_0)$ is the renormalized Fermi velocity and the proximity pairing potential $\Delta\approx g\Delta_0/(g+\Delta_0)$ with $\Delta_0$ as the bulk superconducting pairing gap and $g$ as an effective coupling strength. Note that the external magnetic field $\bm{A}$ does not appear here because it only has $\hat{\bm{x}}$ component in our gauge choice.
The trial wavefunction can be written as $(u_\gamma, v_\gamma)^T e^{i k y}$, and the eigen-equation reads
\begin{equation}
\left(\begin{array}{cc}
  \sigma_{\gamma} \hbar v_{s} k-\mu^{\prime} & \Delta e^{i \sigma_{\gamma}\frac{\varphi }{2} } \\
\Delta e^{-i \sigma_{\gamma}\frac{\varphi }{2} } & \sigma_{\gamma} \hbar v_s k+\mu^{\prime}
\end{array}\right)\left(\begin{array}{l}
  u_\gamma \\
  v_\gamma
\end{array}\right)=\epsilon\left(\begin{array}{l}
  u_\gamma \\
  v_\gamma
\end{array}\right) .
\end{equation}

It is straightforward to obtain the scattering states in the superconducting region, which are fermionic modes in which the electron and hole states are hybridized and propagate chirally along the quantum anomalous Hall–SC interface, the so-called chiral Andreev edge states (CAESs) as:
\begin{equation}
\begin{gathered}
\Psi^S_{e \gamma}=\left(\begin{array}{c}
e^{i \sigma_{\gamma}\frac{\varphi }{2}} \\
-e^{-\beta}
\end{array}\right) e^{i \sigma_{\gamma} k_{s,e} y} \\
\Psi^S_{ h\gamma}=\left(\begin{array}{c}
e^{-\beta} \\
e^{-i \sigma_{\gamma}\frac{\varphi }{2}}
\end{array}\right) e^{i \sigma_{\gamma} k_{s, h} y} \label{Eq_S6}
\end{gathered}
\end{equation}
with definitions $k_{s, e/h}=(\epsilon\pm\sqrt{\Delta^2+\mu^{\prime 2}})/\hbar v_s$, $\beta=\operatorname{arcsinh} \frac{\mu^{\prime}}{\Delta}$.

The total wavefunction thus can be written as
\begin{equation}
  \begin{aligned}
    &\Psi=\left\{\begin{array}{l}
    a_\gamma\left(\begin{array}{c}
    e^{i \sigma_{\gamma} \frac{\varphi}{2}} \\
    -e^{-\beta}
    \end{array}\right) e^{i \sigma_{\gamma} k_{s, e}y}+b_\gamma\left(\begin{array}{c}
    e^{-\beta} \\
    e^{-i \sigma_{\gamma} \frac{\varphi}{2} }
    \end{array}\right) e^{i \sigma_{\gamma} k_{s, h}y}, \quad \gamma=L/R, \quad y \in\left[-\frac{W}{2}, \frac{W}{2} \right] \\
    c_{e\alpha}\left(\begin{array}{l}
    1 \\
    0
    \end{array}\right) e^{i \alpha \left(k_{F, e}+\frac{\pi B W }{2\Phi_0}\right)x}+c_{h\alpha}\left(\begin{array}{l}
    0 \\
    1
    \end{array}\right) e^{i \alpha \left(k_{F, h}-\frac{\pi B W }{2\Phi_0}\right)x}, \quad \alpha=T/B, \quad x \in\left[-\frac{d}{2}, \frac{d}{2} \right]
    \end{array}\right.
    \end{aligned}
  \end{equation}
And to be convenient, we define
\begin{equation}
\begin{array}{r}
c_{e\alpha}(x)=c_{e\alpha} e^{i \alpha \left(k_{F, e}+\frac{\pi B W }{2\Phi_0}\right)x}, c_{h\alpha}(x)=c_{h\alpha} e^{i \alpha \left(k_{F, h}-\frac{\pi B W }{2\Phi_0}\right)x}, \\
a_\gamma(y)=a_\gamma e^{i \sigma_{\gamma} k_{s,e} y}, b_\gamma(y)=b_\gamma e^{i \sigma_{\gamma} k_{s,h} y},
\end{array}
\end{equation}
and
\begin{equation}
  \begin{array}{r}
  c_{e/h}(1)=c_{e/h, T}(-d/2), c_{e/h}(2)=c_{e/h, T}(d/2), c_{e/h}^{\prime}(3)=c_{e/h, B}(d/2), c_{e/h}^{\prime}(4)=c_{e/h, B}(-d/2), \\
  a(1)=a_L(W/2),b(1)=b_L(W/2); a(4)=a_L(-W/2),b(4)=b_L(-W/2);\\
   a^{\prime}(2)=a_R(W/2), b^{\prime}(2)=b_R(W/2);a^{\prime}(3)=a_R(-W/2),b^{\prime}(3)=b_R(-W/2).
  \end{array}
  \end{equation}

We first consider the scattering process between the lower and upper edge through the left-hand side edge.
\begin{equation}
\left(\begin{array}{l}a\left(1\right) \\ b\left(1\right)\end{array}\right)=\left(\begin{array}{cc}e^{i k_{s, e}W} & 0 \\ 0 & e^{i k_{s, h}W}\end{array}\right)\left(\begin{array}{l}a\left(4\right) \\ b\left(4\right)\end{array}\right),
\end{equation}
Using the continuity of the wavefunction, we obtain
\begin{equation}
\begin{aligned}
\left(\begin{array}{l}
c_e^{\prime}\left(4\right) \\
c_h^{\prime}\left(4\right)
\end{array}\right) &=\left(\begin{array}{cc}
e^{i \frac{\varphi}{2}} & e^{-\beta} \\
-e^{-\beta} & e^{-i \frac{\varphi}{2}}
\end{array}\right)\left(\begin{array}{l}
a\left(4\right) \\
b\left(4\right)
\end{array}\right) \\
\left(\begin{array}{l}
c_e\left(1\right) \\
c_h\left(1\right)
\end{array}\right) &=\left(\begin{array}{cc}
e^{i \frac{\varphi}{2}} & e^{-\beta} \\
-e^{-\beta} & e^{-i \frac{\varphi}{2}}
\end{array}\right)\left(\begin{array}{l}
a\left(1\right) \\
b\left(1\right)
\end{array}\right)
\end{aligned}
\end{equation}
Using Eq.~(S10) and (S11), we find
\begin{equation}
\left(\begin{array}{l}
c_e\left(1\right) \\
c_h\left(1\right)
\end{array}\right)=\left(\begin{array}{cc}
e^{i \frac{\varphi}{2}} & e^{-\beta} \\
-e^{-\beta} & e^{-i \frac{\varphi}{2}}
\end{array}\right)\left(\begin{array}{cc}
e^{i k_{s, e}W} & 0 \\
0 & e^{i k_{s, h}W}
\end{array}\right)\left(\begin{array}{cc}
e^{i \frac{\varphi}{2}} & e^{-\beta} \\
-e^{-\beta} & e^{-i \frac{\varphi}{2}}
\end{array}\right)^{-1}\left(\begin{array}{l}
c_e^{\prime}\left(4\right) \\
c_h^{\prime}\left(4\right)
\end{array}\right) .
\end{equation}
We define
\begin{equation}
\left(\begin{array}{c}
c_e\left(1\right) \\
c_h\left(1\right)
\end{array}\right)=S_{41}\left(\begin{array}{c}
c_e^{\prime}\left(4\right) \\
c_h^{\prime}\left(4\right)
\end{array}\right)
\end{equation}
The scattering matrix
\begin{equation}
\begin{aligned}
&S_{41}=\left(\begin{array}{cc}
e^{i \frac{\varphi}{2}} & e^{-\beta} \\
-e^{-\beta} & e^{-i \frac{\varphi}{2}}
\end{array}\right)\left(\begin{array}{cc}
e^{i k_{s, e}W} & 0 \\
0 & e^{i k_{s, h}W}
\end{array}\right)\left(\begin{array}{cc}
e^{i \frac{\varphi}{2}} & e^{-\beta} \\
-e^{-\beta} & e^{-i \frac{\varphi}{2}}
\end{array}\right)^{-1}, \\
&=\frac{e^{i \frac{\epsilon W}{\hbar v_s}}}{\cosh (\beta)}\left(\begin{array}{cc}
\sinh \beta e^{i \delta k W}+e^{-\beta} \cos (\delta k W) & -i \sin (\delta k W) e^{i \frac{\varphi}{2}} \\
-i \sin (\delta k W) e^{-i \frac{\varphi}{2}} & \sinh \beta e^{-i \delta k W}+e^{-\beta} \cos (\delta k W)
\end{array}\right),
\end{aligned}
\end{equation}
where $\delta k=\frac{\sqrt{\Delta^2+\mu'^2}}{\hbar v_s}$. According to $S_{41}$, the Andreev reflection amplitude and normal reflection amplitude are
\begin{equation}
\begin{aligned}
&\left|R_A\right|^2=\frac{\sin ^2(\delta k W)}{\cosh ^2 \beta}=\frac{\sin ^2(\delta k W)}{1+\left(\frac{\mu^{\prime}}{\Delta}\right)^2}, \\
&\left|R_N\right|^2=\frac{\cos ^2(\delta k W)+\sinh ^2 \beta}{\cosh ^2 \beta}=\frac{\cos ^2(\delta k W)+\left(\frac{\mu^{\prime}}{\Delta}\right)^2}{1+\left(\frac{\mu^{\prime}}{\Delta}\right)^2} .
\end{aligned}
\end{equation}

The following features are worth highlighting: (1) The superconductor is gapped, suppressing single-particle tunnelling. As a result, $S_{41}$ is unitary with $\left|R_A\right|^2+\left|R_N\right|^2=1$; (2) Although the edge states are chiral, $R_A$ can be still finite. It means an incoming electron (hole) entering corner 4 can be scattered into an outgoing hole (electron) exiting at corner 1, and at the same time, a cooper pair enters (exits) the superconductor condensate. This process can be seen as a crossed Andreev reflection; (3) The Andreev reflection probability $\left|R_A\right|^2$ exhibits a period oscillation as a function of $\delta k W$, due to the interference between the electron-like and hole-like CAESs after they propagated along their path and accumulating phase difference; (4) The Andreev reflection is suppressed by the increase of $\mu^{\prime} / \Delta$ as the wavefunctions of states behave closer to the wavefunctions of normal chiral edge states [see Eq.~(\ref{Eq_S6})]. The Andreev reflection would be small and finally neglectable when $\mu^{\prime} / \Delta \gg 1$ as $\left|R_A\right|^2 \rightarrow 0$.

Now we define $\psi_{\text{in }}=[c_{e}^{\prime}\left(4\right), c_{h}^{\prime}\left(4\right), c_{e}\left(2\right), c_{h}\left(2\right)]^T$, $\psi_{\text {out }}=[c_{e}\left(1\right), c_{h}\left(1\right),c_{e}^{\prime}\left(3\right), c_{h}^{\prime}\left(3\right)]^T$, and $\psi_{\text {in}}=\mathcal{S}_{N} \psi_{\text {out}}$, $\psi_{\text {out}}=\mathcal{S}_{A} \psi_{\text {in}}$, in order to calculate the supercurrent of the Josephson junction. According to the definitions, the normal transmission matrix (from corner 1 to corner 2 and corner 3 to corner 4 in Fig.~\ref{fig:figS1}).
  \begin{equation}
    \mathcal{S}_{N}=\left(\begin{array}{cccc}
    0 & 0 & e^{i \left(k_{F, e}d+\frac{\pi \Phi }{2\Phi_0}\right)} & 0 \\
    0 & 0 & 0 & e^{i \left(k_{F, h}d-\frac{\pi \Phi }{2\Phi_0}\right)} \\
    e^{i \left(k_{F, e}d+\frac{\pi \Phi }{2\Phi_0}\right)} & 0 & 0 & 0 \\
    0 & e^{i \left(k_{F, h}d-\frac{\pi \Phi }{2\Phi_0}\right)} & 0 & 0
    \end{array}\right)
    \end{equation}
with $\Phi=BWd$ and we define
the scattering at the corners
\begin{equation}
\mathcal{S}_{0}=\left(\begin{array}{cccc}
e^{i \frac{\varphi}{2}} & e^{-\beta}  & 0 & 0 \\
-e^{-\beta} & e^{-i \frac{\varphi}{2}} & 0 & 0 \\
0 & 0 & e^{-i \frac{\varphi}{2}} & e^{-\beta} \\
0 & 0 & -e^{-\beta} & e^{i \frac{\varphi}{2}}
\end{array}\right)
\end{equation}
and the transmission matrix along the left and right edges
\begin{equation}
\mathcal{T}_{s}=\left(\begin{array}{cccc}
e^{i k_{s, e}W} & 0 & 0 & 0 \\
0 & e^{i k_{s, h}W} & 0 & 0 \\
0 & 0 & e^{i k_{s, e}W} & 0 \\
0 & 0 & 0 & e^{i k_{s, h}W}.
\end{array}\right)
\end{equation}
to have
  \begin{equation}
    \mathcal{S}_{A0}=\mathcal{S}_0 \mathcal{T}_s \mathcal{S}_0^{-1}=\left(\begin{array}{cc}
S_{41} & 0\\
0 & S_{23}
\end{array}\right)_{4\times4}
    \end{equation}
where $S_{23}$ is just the $S_{41}$ with $\varphi \rightarrow -\varphi$. Here, we define the $S_{A0}$ as the $S_{A}$ only includes crossed Andreev reflection processes. We will obtain the supercurrent through the Josephson junction using the scattering matrix $S_{A0}$ and $S_N$, as we would show in Sec. C later.

\subsection{B. Scattering matrices in the short junction case}

\begin{figure}
  \centering
  \includegraphics[width=0.9\linewidth]{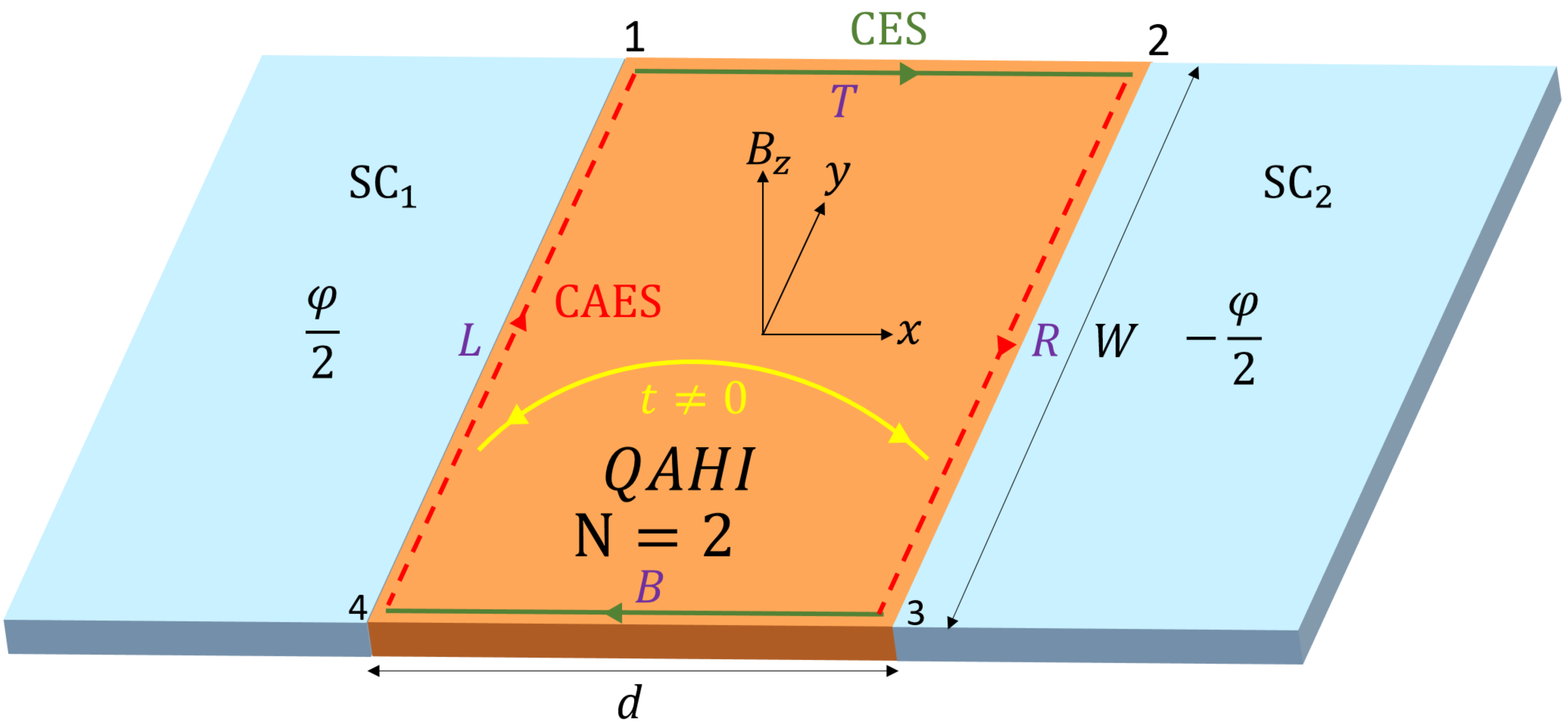}
  \caption{A schematic illustration of
the case where the edge states from the left and right edges hy-
bridize so that the hopping $t\neq 0$. Abbreviations: QAHI (quantum anomalous Hall insulator), CES (chiral edge state), CAES (chiral andreev edge state), SC (superconductor).}
  \label{fig:figS2}
\end{figure}

In this case, we assume there is a small overlap between the wave functions of the CAESs on the left and right edges (as illustrated in Fig.~\ref{fig:figS2}), which induces a small non-chiral component in each original CAES. We rewrite the two scattering modes as \cite{zhang2017quantum}
\begin{equation}
\begin{split}
    \Psi^S_{\alpha +}&=[ r_{\alpha,LL}\zeta_{\alpha L}+t_{\alpha,LR}\zeta_{\alpha  R}]  e^{i  k_{s,\alpha,+} y},\\
    \Psi^S_{\alpha -}&=[ r_{\alpha,RR}\zeta_{\alpha R}+t_{\alpha,RL}\zeta_{\alpha L}]  e^{i  k_{s,\alpha,-} y},
\end{split}
\end{equation}
where $\pm$ labels the propagating direction, $\alpha=e/h$, $\zeta_{e\gamma}=(e^{i \sigma_\gamma \varphi/2}, -e^{-\beta})_\gamma^T$, $\zeta_{h\gamma}=(e^{-\beta}, e^{-i \sigma_\gamma \varphi/2})_\gamma^T$, with $\gamma=L/R$. The overlapping between left and right edges is characterized by
\begin{equation}
t_{\alpha, LR(RL)}=t(d)e^{\mp i \frac{\sigma_\alpha q}{\hbar} \int_{L}^{R} \mathbf{A}\left(y\right) \cdot \hat{\bm{x}} d x}= e^{-\frac{d}{\xi_d}\mp i \sigma_\alpha \pi \frac{\Phi}{\Phi_0}\frac{y}{W}},
\end{equation}
 which is dressed by Peierls substitution and is exponentially suppressed with respect to the junction distance $d$. The coefficient $t(d)=e^{-d/\xi_d}$ describes the hybridization between the left and right edges. We recall that the magnetic flux $\Phi=BdW$ with $W$ as the junction width, $\sigma_{\text{e/h}}=\pm 1$, the quantum flux $\Phi_0=h/2e$, and we denote $\xi_d$ as the decay length of the CAES that should be much smaller than $\xi_s$ because of the large weak link bulk gap, as the consequence, $t(d)\ll1$. Note that the unitary condition requires $|r_{\gamma\gamma}|^2+|t_{\alpha\gamma\gamma'}|^2=1$. And we assume the spectrum of the scattering modes does not change as an approximation, i.e., $k_{s, \alpha, \pm}=\pm\left(\epsilon+\sigma_\alpha\sqrt{\Delta^2+\mu^{\prime 2}}\right)/ \hbar v_s$. Note that the unitary condition requires $|r_{\alpha,\gamma\gamma}|^2+|t_{\alpha,\gamma\gamma'}|^2=1$, which permits an arbitrary phase difference between $r_{\alpha,\gamma\gamma}$ and $t_{\alpha,\gamma\gamma'}$. Here we choose $r_{\alpha,\gamma\gamma}=r(d)=\sqrt{1-\text{exp}(-2d/\xi_d)}$ to preserve the inversion symmetry.

The total wavefunction propagating along the y-direction (forward and backward) can be expanded with the scattering modes $\Psi_{\alpha,+}^{S}$ and $\Psi_{\alpha,-}^{S}$, which   can be written as:
\begin{equation}
\Psi_+=a[r\left(\begin{array}{c}
  e^{i\frac{\varphi}{2}} \\
-e^{-\beta}
\end{array}\right)_L +t e^{-i \pi \frac{\Phi}{\Phi_0}\frac{y}{W}}\left(\begin{array}{c}
e^{-i\frac{\varphi}{2}} \\
-e^{-\beta}
\end{array}\right)_R ]e^{i k_{F1}y}+
b[r\left(\begin{array}{c}
  e^{-\beta} \\
  e^{-i\frac{\varphi}{2}}
\end{array}\right)_L +t e^{+i \pi \frac{\Phi}{\Phi_0}\frac{y}{W}}\left(\begin{array}{c}
e^{-\beta} \\
  e^{i\frac{\varphi}{2}}
\end{array}\right)_R ]e^{i k_{F2}y}
\end{equation}
\begin{equation}
\Psi_-=a'[r\left(\begin{array}{c}
  e^{-i\frac{\varphi}{2}} \\
-e^{-\beta}
\end{array}\right)_R +t e^{+i \pi \frac{\Phi}{\Phi_0}\frac{y}{W}}\left(\begin{array}{c}
e^{i\frac{\varphi}{2}} \\
-e^{-\beta}
\end{array}\right)_L ]e^{-i k_{F1}y}+
b'[r\left(\begin{array}{c}
  e^{-\beta} \\
  e^{i\frac{\varphi}{2}}
\end{array}\right)_R +t e^{-i \pi \frac{\Phi}{\Phi_0}\frac{y}{W}}\left(\begin{array}{c}
e^{-\beta} \\
  e^{-i\frac{\varphi}{2}}
\end{array}\right)_L]e^{-i k_{F2}y}
\end{equation}
Here, for the compact of notations,  we have used $k_{F1}=\left(\epsilon+\sqrt{\Delta^2+\mu^{\prime 2}}\right)/\hbar v_s$, $k_{F2}=\left(\epsilon-\sqrt{\Delta^2+\mu^{\prime 2}}\right)/ \hbar v_s$ to replace above $k_{s, \alpha, \pm}$. And $r$, $t$ represent $r(d)$, $t(d)$. By matching the boundary conditions, we obtain the following:
\begin{equation}
\left(\begin{array}{c}
c_e(1) \\
c_h(1)
\end{array}\right)=r\left(
                     \begin{array}{cc}
                       e^{i\frac{\varphi+k_{F1}W}{2}} & e^{-\beta+i\frac{ k_{F2}W}{2}} \\
                       -e^{-\beta+i\frac{k_{F1}W}{2}} & e^{-i\frac{\varphi-k_{F2}W}{2}} \\
                     \end{array}
                   \right)
                  \left(\begin{array}{c}
a \\
b
\end{array}\right)+t\left(
                     \begin{array}{cc}
                       e^{i\frac{\varphi-k_{F1}W}{2}+\frac{i\pi \Phi}{2\Phi_0}} & e^{-\beta-i\frac{k_{F2}W}{2}-\frac{i\pi \Phi}{2\Phi_0}} \\
                       -e^{-\beta-i\frac{k_{F1}W}{2}+\frac{i\pi \Phi}{2\Phi_0}} & e^{-i\frac{\varphi+k_{F2}W}{2}-\frac{i\pi \Phi}{2\Phi_0}} \\
                     \end{array}
                   \right)
\left(\begin{array}{c}
a' \\
b'
\end{array}\right)\label{Eq24}
\end{equation}

\begin{equation}
\left(\begin{array}{c}
c_e^\prime(4) \\
c_h^\prime(4)
\end{array}\right)=r\left(
                     \begin{array}{cc}
                       e^{i\frac{\varphi-k_{F1}W}{2}} & e^{-\beta-i\frac{k_{F2}W}{2}} \\
                       -e^{-\beta-i\frac{k_{F1}W}{2}} & e^{-i\frac{\varphi+k_{F2}W}{2}} \\
                     \end{array}
                   \right)
                  \left(\begin{array}{c}
a \\
b
\end{array}\right)+t\left(
                     \begin{array}{cc}
                       e^{i\frac{\varphi+k_{F1}W}{2}-\frac{i\pi \Phi}{2\Phi_0}} & e^{-\beta+i\frac{k_{F2}W}{2}+\frac{i\pi \Phi}{2\Phi_0}} \\
                       -e^{-\beta+i\frac{k_{F1}W}{2}-\frac{i\pi \Phi}{2\Phi_0}} & e^{-i\frac{\varphi-k_{F2}W}{2}+\frac{i\pi \Phi}{2\Phi_0}} \\
                     \end{array}
                   \right)
\left(\begin{array}{c}
a' \\
b'
\end{array}\right)
\end{equation}

Similarly,
\begin{equation}
\left(\begin{array}{c}
c_e(2) \\
c_h(2)
\end{array}\right)=r\left(
                     \begin{array}{cc}
                       e^{-i\frac{\varphi+k_{F1}W}{2}} & e^{-\beta-i\frac{k_{F2}W}{2}} \\
                       -e^{-\beta-i\frac{k_{F1}W}{2}} & e^{i\frac{\varphi-k_{F2}W}{2}} \\
                     \end{array}
                   \right)
                  \left(\begin{array}{c}
a' \\
b'
\end{array}\right)+t\left(
                     \begin{array}{cc}
                       e^{-i\frac{\varphi-k_{F1}W}{2}-\frac{i\pi \Phi}{2\Phi_0}} & e^{-\beta+i\frac{k_{F2}W}{2}+\frac{i\pi \Phi}{2\Phi_0}} \\
                       -e^{-\beta+i\frac{k_{F1}W}{2}-\frac{i\pi \Phi}{2\Phi_0}} & e^{i\frac{\varphi+k_{F2}W}{2}+\frac{i\pi \Phi}{2\Phi_0}} \\
                     \end{array}
                   \right)
\left(\begin{array}{c}
a \\
b
\end{array}\right)
\end{equation}

\begin{equation}
\left(\begin{array}{c}
c_e^\prime(3) \\
c_h^\prime(3)
\end{array}\right)=r\left(
                     \begin{array}{cc}
                       e^{-i\frac{\varphi-k_{F1}W}{2}} & e^{-\beta+i\frac{k_{F2}W}{2}} \\
                       -e^{-\beta+i\frac{k_{F1}W}{2}} & e^{i\frac{\varphi+k_{F2}W}{2}} \\
                     \end{array}
                   \right)
                  \left(\begin{array}{c}
a' \\
b'
\end{array}\right)+t\left(
                     \begin{array}{cc}
                       e^{-i\frac{\varphi+k_{F1}W}{2}+\frac{i\pi \Phi}{2\Phi_0}} & e^{-\beta-i\frac{k_{F2}W}{2}-\frac{i\pi \Phi}{2\Phi_0}} \\
                       -e^{-\beta-i\frac{k_{F1}W}{2}+\frac{i\pi \Phi}{2\Phi_0}} & e^{i\frac{\varphi-k_{F2}W}{2}-\frac{i\pi \Phi}{2\Phi_0}} \\
                     \end{array}
                   \right)
\left(\begin{array}{c}
a \\
b
\end{array}\right) \label{Eq27}
\end{equation}
After some massages, we can rewrite Eq.~\ref{Eq24} to \ref{Eq27} as new matrix equations \begin{equation}
    \psi_{\text {in }}=\mathcal{S}_0(r\mathcal{T}_{0}^{-1}+t\mathcal{T}_{0}\mathcal{U}_{0}) \psi, \psi_{\text {out }}=\mathcal{S}_0(r\mathcal{T}_{0}+t\mathcal{T}_{0}^{-1}\mathcal{U}_{0}^{*}) \psi, \label{Eq_S28}
\end{equation}
where the incoming state $\psi_{in}$, outgoing state $\psi_{out}$, and the scattering matrix at the corners $S_0$ have the same definitions  in Sec. A, and for the sake of convenience, we have defined
\begin{eqnarray}
&&\psi=(a, b, a', b')^T\\
&&\mathcal{T}_{0}=\left(\begin{array}{cccc}
e^{i\frac{ k_{F1}W}{2}} & 0 & 0 & 0 \\
0 & e^{i \frac{ k_{F2}W}{2}} & 0 & 0 \\
0 & 0 & e^{i\frac{ k_{F1}W}{2}} & 0 \\
0 & 0 & 0 & e^{i\frac{ k_{F2}W}{2}}
\end{array}\right),\\
&&\mathcal{U}_{0}=\left(\begin{array}{cccc}
0 & 0 & e^{ -\frac{i\pi\Phi}{2\Phi_0}} & 0 \\
0 & 0 & 0 & e^{ \frac{i\pi\Phi}{2\Phi_0}} \\
e^{- \frac{i\pi\Phi}{2\Phi_0}} & 0 & 0 & 0 \\
0 & e^{ \frac{i\pi\Phi}{2\Phi_0}} & 0 & 0
\end{array}\right).
\end{eqnarray}

By eliminating $\psi$ in Eq.~\ref{Eq_S28}, we obtain
$\psi_{out}=\mathcal{S}_A\psi_{in}$, where the  scattering matrix $\mathcal{S}_A$ indicates how an incoming state $\psi_{in}$ is scattered into an outgoing state $\Psi_{out}$. Being similar to Eq.~(S19), we have \begin{equation}
\mathcal{S}_{A}=\mathcal{S}_0 \mathcal{T}_{s}^{\prime} \mathcal{S}_0^{-1}.
\end{equation}
The transmission matrix $\mathcal{T}_{s}^{\prime}$ is shifted from the one $\mathcal{T}_{s}$ in Sec. A to
\begin{equation}\mathcal{T}_{s}^{\prime}\approx \mathcal{T}_{s} (1-2i\mathcal{U}_{1}t/r),
\end{equation}
 in the $t\ll r$ limit, and where
 \begin{equation}
\mathcal{U}_{1}\equiv\left(\begin{array}{cccc}
0 & 0 & \sin(k_{F1}W-\phi/2) & 0 \\
0 & 0 & 0 &  \sin(k_{F2}W+\phi/2)  \\
 \sin(k_{F1}W-\phi/2)  & 0 & 0 & 0 \\
0 &  \sin(k_{F2}W+\phi/2)  & 0 & 0
\end{array}\right),
\end{equation}
By inserting Eq.~(S33) into Eq.~(S32), we find the  scattering matrix
\begin{equation}
    S_{A} \approx \mathcal{S}_{A0} (1-2i\mathcal{U}_{1}^{\prime}t/r)=\mathcal{S}_{A0}+\delta\mathcal{S}_{A}
\end{equation}
where $\mathcal{U}_{1}^{\prime} \equiv \mathcal{S}_0 \mathcal{U}_{1} \mathcal{S}_0^{-1}$ and the scattering matrix shift due to the overlapping of edge states is given by
\begin{equation}
   \delta\mathcal{S}_{A} =-2iS_{A0}\mathcal{U}_{1}^{\prime}t/r.
\end{equation}
 Finally, it is worth noting that the normal transition matrix
 $\mathcal{S}_{N}$ with $\Psi_{in}=\mathcal{S}_{N}\Psi_{out}$ remains the same as the one in the long junction since the states at the top and bottom edges remain unchanged. In the next subsection, we calculate the Fraunhofer patterns at long and short junction limits using the scattering matrices $S_{A}$ and $S_{N}$ we obtained.

\subsection{C. Calculate the Fraunhofer patterns using the scattering matrices}
Using the formula from \cite{brouwer1997anomalous}, we can get the flux-dependent current-phase relation (CPR) of the JJ:
\begin{equation}
  I_s(\varphi,\Phi)=-\frac{2 e k_B T}{\hbar} \frac{d}{d \varphi} \sum_{n=0}^{\infty} \ln \operatorname{det}\left[1-\mathcal{S}_A\left(i \omega_n, \varphi,\Phi\right) \mathcal{S}_N\left(i \omega_n, \varphi,\Phi\right)\right],
  \end{equation}
  where $\omega_n=(2 n+1)\pi k_B T$ are fermionic Matsubara frequencies, replacing $\epsilon$ in the scattering matrices. At the same time, we can also address the flux-dependent critical supercurrent $I_c(\Phi)=\max_\varphi|I_s(\varphi, \Phi)|$, which presents the periodic Fraunhofer oscillations.

  In the following, we want to show the explicit form of $I_s(\varphi,\Phi)$. For the long junction case, we have
   \begin{equation}
  I_{s0}(\varphi,\Phi)=\sum_{n=0}^{\infty } \frac{2e\omega  \sin\varphi }{\cosh^{2}\beta \csc ^{2}\alpha \cosh \frac{2 \pi^2 (2n+1)\omega}{\omega_0}+\cos\varphi +(\cosh 2\beta-\cosh^{2}\beta\csc ^{2}\alpha)\cos\left ( \phi+2\phi_0 \right ) +\cot \alpha \sin \left ( \phi+2\phi_0 \right )\sinh 2\beta},
  \end{equation}
  where $\omega=k_BT/\hbar$, $\omega_0=\pi \left( \frac{d}{v_{\mathrm{F}}}+\frac{W}{v_{\mathrm{s}}}\right)^{-1}$, $\phi_0=\mu d/\hbar v_F$, $\alpha=\frac{W\sqrt{\Delta^2+\mu^{\prime 2}}}{\hbar v_s}$. To get a simplified form, at the condition $2 \pi^2 \omega / \omega_0 \gg 1$, the first term in the denominator dominates, and the $n=0$ term dominates in the Matsubara frequency summation. If we set $\mu^\prime=0$ meaning $\beta=0$, then we obtain
   \begin{align}
  I_{s0}(\varphi,\Phi) &\approx  \frac{2e\omega  \sin\varphi }{ \csc ^{2}\alpha \cosh \frac{2 \pi^2 \omega}{\omega_0}+\cos\varphi -\cot ^{2}\alpha\cos\left ( \phi+2\phi_0 \right ) }\\
  & \approx \frac{2e\omega \sin ^{2}\alpha \sin\varphi }{  \cosh \frac{2 \pi^2 \omega}{\omega_0}}\left [ 1-\frac{\sin ^{2}\alpha \cos\varphi}{ \cosh \frac{2 \pi^2 \omega}{\omega_0}} +\frac{\cos ^{2}\alpha}{ \cosh \frac{2 \pi^2 \omega}{\omega_0}} \cos\left ( \phi+2\phi_0 \right )  \right ]
  \end{align}

  For the short junction case with the correction from coupling, we use the approximate form of the S-matrix,
  \begin{align}
  I_s(\varphi,\Phi)& \approx -\frac{2 e k_B T}{\hbar} \frac{d}{d \varphi} \sum_{\omega_n} \ln \operatorname{det}\left[ 1-\left( \mathcal{S}_{A0}+\delta\mathcal{S}_{A}\right) \mathcal{S}_N +\mathcal{S}_{A0}\mathcal{S}_N\delta\mathcal{S}_{A}\mathcal{S}_N \right]\\
  & = -\frac{2 e k_B T}{\hbar} \frac{d}{d \varphi} \sum_{\omega_n} \ln \operatorname{det}\left( 1-\mathcal{S}_{A0} \mathcal{S}_N \right)-\frac{2 e k_B T}{\hbar} \frac{d}{d \varphi} \sum_{\omega_n} \ln \operatorname{det}\left( 1-\delta\mathcal{S}_{A}\mathcal{S}_N \right)\\
  & = I_{s0}(\varphi,\Phi)+\delta I_{s}(\varphi,\Phi),
  \end{align}
  in which $I_{s0}$ is what we have calculated above, and $\delta I_{s}$ reads
 \begin{align}
 \delta I_{s}(\varphi,\Phi) & \approx -\frac{2 e k_B T}{\hbar} \frac{d}{d \varphi} \sum_{\omega_n} \ln\left[ 1- \operatorname{Tr} \left(\delta\mathcal{S}_{A}\mathcal{S}_N\right) \right]\\
 & \approx \frac{2 e k_B T}{\hbar} \sum_{\omega_n} \frac{d}{d \varphi} \operatorname{Tr} \left(\delta\mathcal{S}_{A}\mathcal{S}_N\right)\\
 & = \frac{4 e \omega t}{r} \sum_{n=0}^{\infty} (1+\tanh \beta ) e^{-\frac{(2n+1)\pi \omega d}{ v_F}} \left [ e^{-\frac{(2n+1)2\pi \omega W}{ v_s} } \cos\left (  2\alpha +\phi_0 \right )  -\cos \left ( \phi+\phi_0 \right )    \right ] \sin \varphi \\
 & \approx -\frac{4 e \omega t}{r} \sum_{n=0}^{\infty} (1+\tanh \beta ) e^{-\frac{(2n+1)\pi \omega d}{ v_F}} \cos \left ( \phi+\phi_0 \right ) \sin \varphi .
\end{align}
With the same spirit, we simplify this formula and obtain
\begin{equation}
    \delta I_{s}(\varphi,\Phi)  \approx -4 e \omega t  e^{-\frac{\pi \omega d}{ v_F}} \cos \left ( \phi+\phi_0 \right ) \sin \varphi .
\end{equation}

  Finally, we can summarize a general formula for both the long and short junction
  \begin{equation}
    I_{s}\left(\varphi,\Phi\right) \approx \left[ I_{0}+I_{1} \cos \left( \phi+ 2\phi_0 \right)+I_{2} \cos \left(  \phi + \phi_0 \right) \right] \sin \varphi
  \end{equation}
with $ I_0=2 e \omega    \sin ^2 \alpha \, \mathrm{sech\,} \frac{2 \pi^2 \omega}{\omega_0}$, $I_1=2 e \omega  \sin ^2 \alpha \cos ^2 \alpha \, \mathrm{sech\,} ^2\frac{2 \pi^2 \omega}{\omega_0}$, and $I_2=-4 e \omega t  e^{-\frac{\pi \omega d}{ v_F}}$. Here we have neglected the second harmonic contribution of $\varphi$.

\begin{figure}[h]
  \centering
  \includegraphics[width=0.70\linewidth]{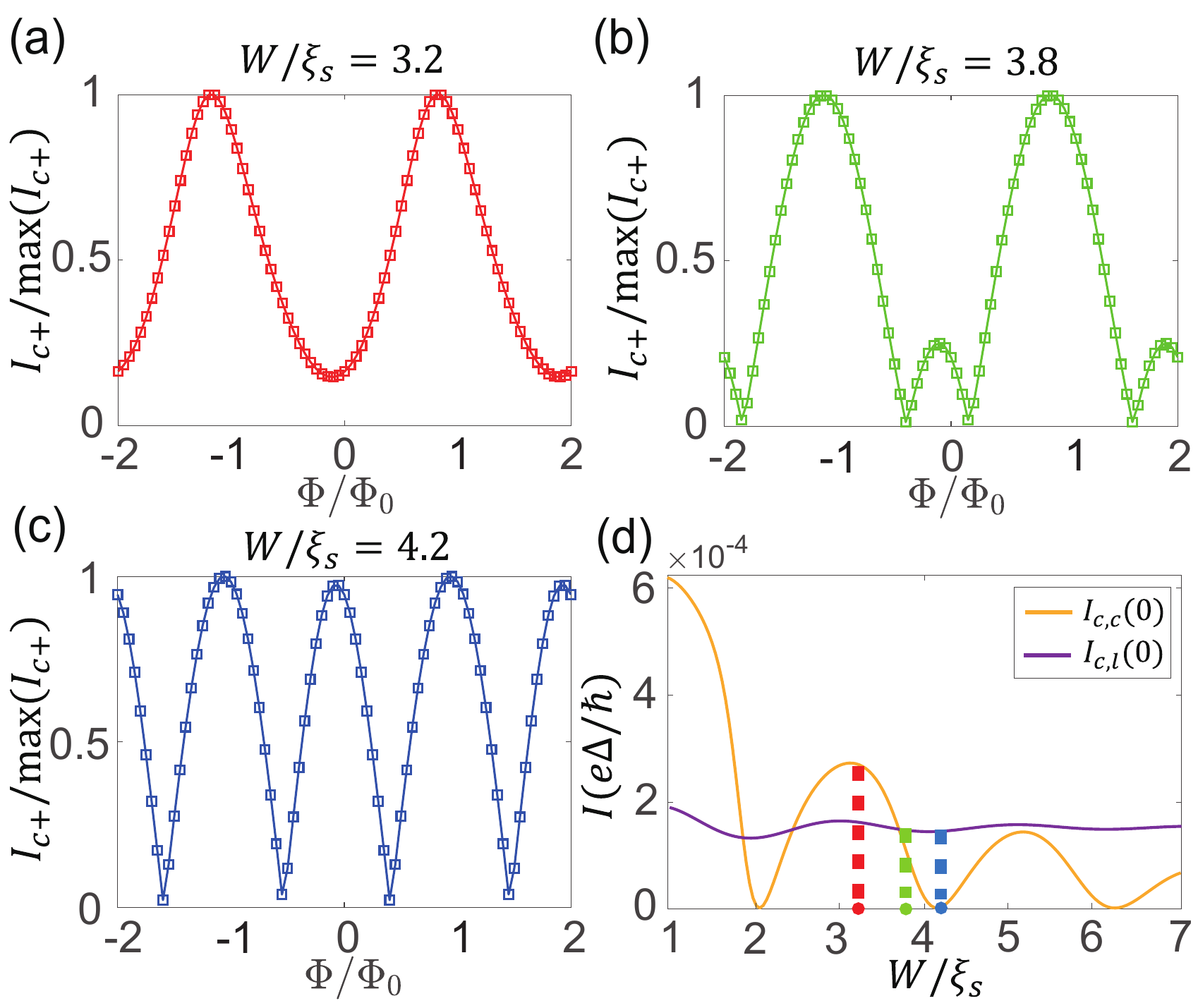}
  \caption{(a), (b), and (c): Fraunhofer patterns for three different junction width $W/\xi_s=3.2$, $W/\xi_s=3.8$ and $W/\xi_s=4.2$, corresponding to the three highlight points in Fig.~\ref{fig:figS3}(d). (d) Junction width dependence of $I_{c,c}$ and $I_{c,l}$ for $d/\xi_s=0.84$. $I_{c,c}$ is the critical supercurrent from local Andreev reflections with zero magnetic flux, and $I_{c,l}$ is the critical supercurrent from local Andreev reflections with zero magnetic flux. In this calculation, we set the parameters as $v_F=2$, $\Delta_0=0.08$, $\mu=0.02$, and $\mu'=0.1$. The coupling strength $g=\Delta_0/5$, which yields $v_s=1.6$ and $\Delta=\Delta_0/6$. The temperature is $k_B T=0.05\Delta_0$. To describe the coupling of the edge states, the parameter $\xi_d$ is set to be 0.1$\xi_s$, where $\xi_s$ is the superconductor coherent length.}
  \label{fig:figS3}
\end{figure}

\subsection{D. Parameter dependences of the Fraunhofer patterns}

 As we have mentioned in the main text, $I_c(\Phi)$ respects $h/e$ ($h/2e$) oscillations if $I_{s,c}$ ($I_{s,l}$) dominates. A concrete way to see when these conditions will be reached is to consider the scaling law of $I_0$ and $I_2$ with respect to the sample size and temperature. From the form of $I_0$, $I_2$, it can be found that both $I_0$ and $I_2$ are suppressed with respect to the length of the junction: $I_0\propto e^{- d/\xi_0}$ with $\xi_0^{-1}= 2\pi k_BT/\hbar v_F$ and $I_2\propto e^{-d/\xi_2}$ with $\xi_2^{-1}\approx \xi_d^{-1}+(2\xi_0)^{-1}$. Note that $I_2$ has a faster suppression (or increase) with respect to $d$ since, in general, $\xi_2 \ll \xi_0$, for $\xi_s/\xi_d \gg 1 \gg \pi k_B T /\Delta_0$. Besides, $I_0$ is also suppressed with respect to the junction width: $I_0 \propto \sin^2(\delta k W) e^{- W v_F/\xi_0 v_S}$, while $I_2$ is independent of $W$, as $I_0$ arises from crossed Andreev reflections and $I_2$ arises from local ones. Furthermore, the temperature dependence of $I_0$ and $I_2$ are different ( $I_0\propto T e^{-2 k_BT \pi (d/\hbar v_F+W/\hbar v_s)}$ and $I_2\propto T e^{-k_B \pi T d/\hbar v_F}$), and apparently, $I_2$ tends to dominate at higher temperatures. According to these observations, three simple ways to drive the $h/e$ and $h/2e$ oscillations crossover are expected: (i) reduce the junction length $d$, which changes $I_2$ more dramatically; (ii) increase the junction width $W$, which suppresses $I_0$; (iii) increase the temperature, which $I_2$ will dominate over $I_0$ at high enough temperature.

Another thing we want to supplement here is the width dependencies of the Fraunhofer patterns. If we fix $d=0.84\xi_s$ and increase $W$, $I_{c,l}$ is not sensitive, but $I_{c,c}$ is globally exponentially suppressed with an addition oscillation due to the $\sin^2(2\delta k W)$ factor. To further justify the role of junction width, we choose three points $W/\xi_s=3.2$, $W/\xi_s=3.8$ and $W/\xi_s=4.2$ in Fig.~\ref{fig:figS3}(d) to plot the Fraunhofer patterns as (a), (b) and (c), respectively. It is clear that the Fraunhofer pattern experiences a crossover from $h/e$ oscillations to $h/2e$ oscillations with the increase of junction width $W$, due to $I_{c,l}$ gradually dominating over $I_{c,c}$ during the increase of junction width $W$. 

\subsection{E. Application to the recent experiments}

\begin{figure}[h]
  \centering
  \includegraphics[width=0.9\linewidth]{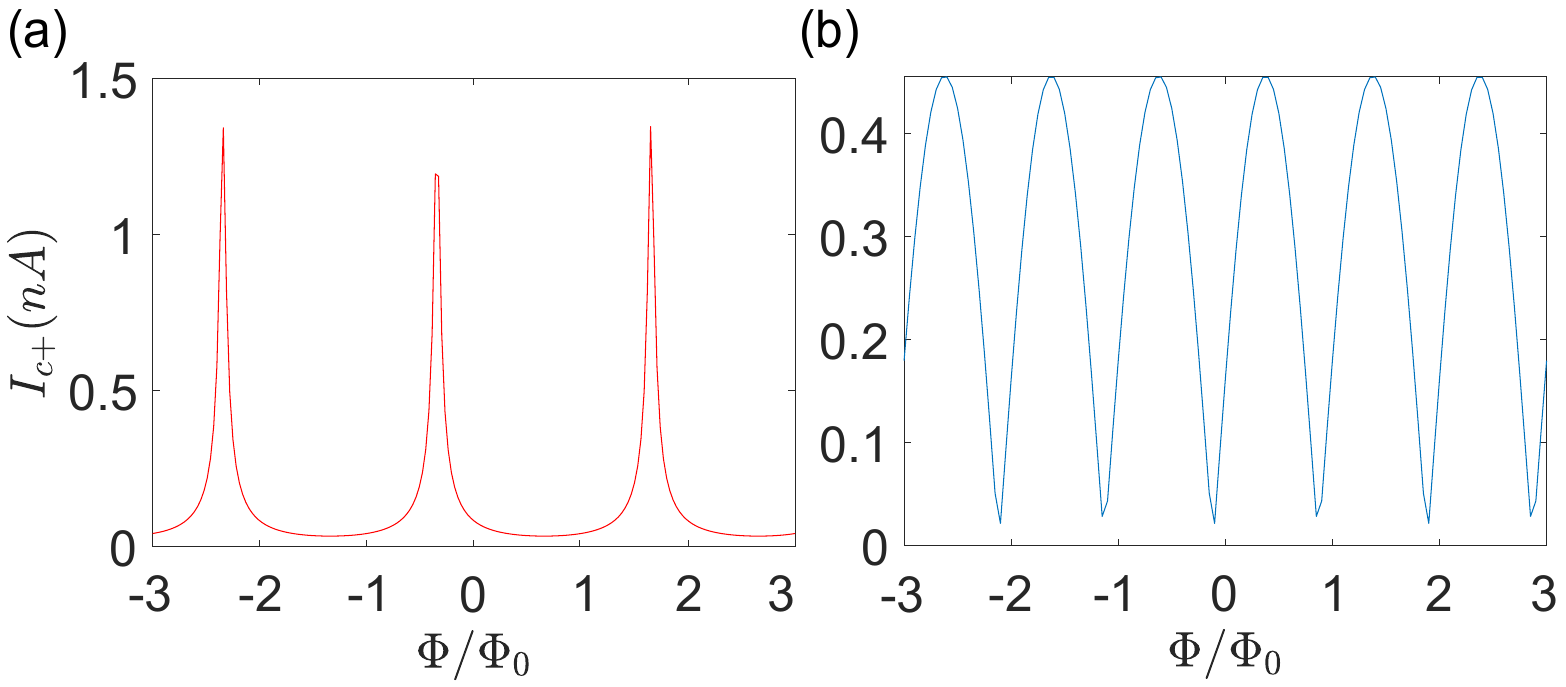}
  \caption{The resulting Fraunhofer patterns from the scattering matrix theory for two groups of different experiment parameters, (a) is for Ref.~\cite{vignaud2023evidence}, which presents the 2$\Phi_0$ period. (b) is for Ref.~\cite{amet2016supercurrent}, which presents the $\Phi_0$ period.}
  \label{fig:figS4}
\end{figure}

Our theory can explain the distinct observations in the recent experiments well. Specifically, the  experiment \cite{vignaud2023evidence} reported the $h/e$ oscillations, being different from Ref.~\cite{amet2016supercurrent}. The key difference we found is that the junction width $W=180$ nm in Ref.~\cite{vignaud2023evidence} is much narrower than the one Ref.~\cite{amet2016supercurrent} where $W=2.4\mu$ m. As a result, the supercurrent $I_{c,c}$ of Ref.~\cite{vignaud2023evidence} has a larger contribution from crossed Andreev reflections. In the experiment of \cite{vignaud2023evidence}, the realistic parameters are: $v_F=1.4\times 10^5m/s$, $v_s=2.1\times 10^4m/s$, $\Delta_0=0.9$ meV, $W=180$ nm, $d=140$ nm, $T=10$ mK. Substituting these values into the scattering matrices and using Eq. S37, we can obtain the critical current as well as Fraunhofer oscillations. We find the calculated supercurrent $I_{c}\approx 1.5$ nA, which is close to the experimental value. Importantly, the period of Fraunhofer oscillations remains 2$\Phi_0$ even though the coupling between the two edges is introduced. This is because, for small $W$, the contribution from $I_{s,c}$ dominates over the $\delta I_{s}$. We also make some comments for the experiment of \cite{amet2016supercurrent}. In that experiment, $v_F=7\times 10^5m/s$, $v_s=4\times 10^4m/s$, $\Delta_0=1.2$ meV, $W=2.4\mu m$, $d=300$ nm, $T=40$ mK. And the supercurrent $I_{c}\approx 0.5$ nA. In this case, The large $W$ makes $I_{s,c}$ very tiny, and $\delta I_{s}$ dominates if a small edge coupling is assumed, giving birth to the $\Phi_0$ Fraunhofer oscillations. We present the detailed calculation results for these two cases in Fig.~\ref{fig:figS4}. 

  \section*{\bf{\uppercase\expandafter{NOTE 2: Numerical results from lattice model}}}

  \subsection{F. The recursive Green's function method}
  In this note, we describe the recursive Green's function method we used to simulate the Josephson junction with details. The barrier part of the junction is chosen as the Qi-Wu-Zhang (QWZ) \cite{qi2006topological} lattice model of QAHI with the basis $[c_{\bm{j},\uparrow}, c_{\bm{j},\downarrow}]^T$ and parameters $(M,A,B)=(1,2,2)$ to make the Hamiltonian in the topological regime:
    \begin{equation}
      \hat{H}= \sum_{\bm{j}}\left(c_{\bm{j}+\hat{\bm{x}}}^{\dagger}c_{\bm{j}} \otimes \frac{i A \hat{\sigma}_x- B \hat{\sigma}_z}{2}+c_{\bm{j}+\hat{\bm{y}}}^{\dagger}c_{\bm{j}} \otimes \frac{i A \hat{\sigma}_y- B \hat{\sigma}_z}{2}+\text { h.c. }\right)
      +M \sum_{\bm{j}} c_{\bm{j}}^{\dagger}c_{\bm{j}} \otimes \hat{\sigma}_z
      \end{equation}

  To simulate two superconducting electrodes, a simple lattice model
  with an s-wave pairing can be used, written as
  \begin{align}
  H_{SC}^{R/L} & =\sum_{\bm{j}_{R/L},\sigma=\uparrow\downarrow}t_{sc}\left(a_{\bm{j}_{R/L}\sigma}^{\dagger}a_{\bm{j}_{R/L}+\hat{\bm{x}},\sigma}+a_{\bm{j}_{R/L}\sigma}^{\dagger}a_{\bm{j}_{R/L}+\hat{\bm{y}},\sigma}+h.c.\right)-\mu_{sc}\\
   & +\frac{1}{2}\sum_{\bm{j}_{R/L}}\left[\Delta_{0}e^{i\varphi_{R/L}}(a_{\bm{j}_{R/L}\uparrow}^{\dagger}a_{\bm{j}_{R/L}\downarrow}^{\dagger}-a_{\bm{j}_{R/L}\downarrow}^{\dagger}a_{\bm{j}_{R/L}\uparrow}^{\dagger})+h.c.\right].
  \end{align}
  Here $a^{\dagger}$ is the electron creation operator in the superconducting region, and $\varphi_{R/L}$ is the pairing phase of the right/left superconducting electrode. The calculated Josephson current is dependent on the pairing
  phase difference between the two superconducting electrodes $\Delta\varphi=\varphi_{R}-\varphi_{L}$.
  The coupling between the superconducting electrodes and the central
  device (QWZ model) is added onto the adjacent sites
  $<\bm{j}_{R/L},\bm{j}>$ between them as
  \begin{equation}
H_{coup}^{R/L}=\sum_{<\bm{j}_{R/L},\bm{j}>}\sum_{\sigma=\uparrow\downarrow}\left(t_{coup}a_{\bm{j}_{R/L}\sigma}^{\dagger}c_{\bm{j},\sigma}+h.c.\right)\label{eq:coupling_Hamil}
  \end{equation}

  When an external field $\bm{B}=B_{0}\hat{\bm{z}}$ is exerted on the
  Josephson junction, as the protection of the pairing against
  the external field is quite weak in the central region, the $\bm{B}$ field can penetrate
  through it without entering the superconducting electrodes. Then we
  can use the Peierls substitution to simulate the effect of the external
  field on the central device, written as
  \begin{equation}
  t_{\bm{jj^{\prime}}}\rightarrow t_{\bm{jj^{\prime}}}\exp\left(i\frac{e}{\hbar}\int_{\bm{j}}^{\bm{j^{\prime}}}\bm{A}(\bm{r})\cdot d\bm{r}\right).
  \end{equation}
  For the sake of comparison with the analytical method, we adopt the Landau gauge as
  $\bm{A}=-yB_{0}\hat{\bm{x}}$.

  With all these preliminary modelling preparation, we can now go on
  with the calculation based on the recursive lattice Green's function
  method \cite{furusaki1994dc,ando1991quantum,asano2001numerical,asano2003josephson,diez2023symmetry,xie2023varphi,hu2023josephson}. We assume the super-current flows along
  the $x$-direction and the length of the central device along $x$-direction
  is $N_{x}$ sites. We label the $j_{x}=1$ site as the left end of
  the central device and $j_{x}=N_{x}$ as the right end.

  In the numerical process, the self-energy $\Sigma_{SC}^{R/L}(i\omega_{n})$
  of the left/right superconducting electrodes onto the central devices
  is first calculated as
  \begin{equation}
  \Sigma_{SC}^{L/R}(i\omega_{n})=V_{1/N_{x},L/R}^{coup}G_{SC}^{L/R}(i\omega_{n})\left(V_{1/N_{x},L/R}^{coup}\right)^{\dagger}.
  \end{equation}
  Here $G_{SC}^{L/R}(i\omega_{n})$ is the Nambu Green's function of
  the left/right superconducting electrode with the Matsubara frequency
  $\omega_{n}=(2n+1)\pi k_{B}T$, which can be calculated iteratively
  by assuming the electrodes are semi-infinite. $V_{1/N_{x},L/R}^{coup}$
  is the coupling matrix between the left/right end of the central device
  and the corresponding superconducting electrode, the form of which
  can be gotten from $H_{coup}^{R/L}$ in Eq.~\ref{eq:coupling_Hamil}.

  Next, we start from the Nambu Green's function of the two ends of the
  central devices
  \begin{align}
  G_{11}^{L}(i\omega_{n}) & =\left[i\omega_{n}-H_{11}^{isol}-\Sigma_{SC}^{L}(i\omega_{n})\right]^{-1},\\
  G_{N_{x}N_{x}}^{R}(i\omega_{n}) & =\left[i\omega_{n}-H_{N_{x}N_{x}}^{isol}-\Sigma_{SC}^{R}(i\omega_{n})\right]^{-1}.
  \end{align}
  Here $H_{xx}^{isol}$ with $x=1,2,...,N_{x}$ represent the BdG Hamiltonian
  of an isolated column (or slice) at $j_{x}=x$ of the central device.
  Explicitly in our case,
  \[
  H_{xx}^{isol}=\left(\begin{array}{cc}
  H_{xx}^{ee}\\
   & H_{xx}^{hh}
  \end{array}\right)
  \]
  with $H_{xx}^{ee}=\sum_{j_{y}}\left[c_{\bm{j}}^{\dagger}\hat{h}_{0}^{j_{y}}c_{\bm{j}}+\left(c_{\bm{j}}^{\dagger}\hat{t}_{y}c_{\bm{j}+\hat{\bm{y}}}+h.c.\right)\right]$,
  $H_{xx}^{hh}=-(H_{xx}^{ee})^{*}$. Here $\bm{j}=(x,j_{y})$.

  Starting from both ends, the Nambu Green's function of the columns
  (or slices) inside the central device can be calculated recursively
  by projecting Dyson's equation between adjacent columns
  \begin{align}
  \Sigma_{xx}^{L}(i\omega_{n}) & =V_{x+1,x}G_{xx}^{L}(i\omega_{n})V_{x+1,x}^{\dagger}\\
  \Sigma_{xx}^{R}(i\omega_{n}) & =V_{x-1,x}G_{xx}^{R}(i\omega_{n})V_{x-1,x}^{\dagger}\\
  G_{x+1,x+1}^{L}(i\omega_{n}) & =\left[i\omega_{n}-H_{x+1,x+1}^{isol}-\Sigma_{xx}^{L}(i\omega_{n})\right]^{-1}\\
  G_{x-1,x-1}^{R}(i\omega_{n}) & =\left[i\omega_{n}-H_{x-1,x-1}^{isol}-\Sigma_{xx}^{R}(i\omega_{n})\right]^{-1}.
  \end{align}
  Explicitly in our case
  \[
  V_{x+1,x}^{\dagger}=V_{x-1,x}=\left(\begin{array}{cc}
  V_{x}^{ee}\\
   & V_{x}^{hh}
  \end{array}\right)
  \]
  where $V_{x}^{ee}=\sum_{j_{y}}c_{\bm{j}}^{\dagger}\hat{t}_{x}c_{\bm{j}+\hat{\bm{x}}}$
  and $V_{x}^{hh}=-\left(V_{x}^{ee}\right)^{*}$. Note that the superscript
  $R/L$ means that the Nambu Green's function $G^{R/L}$ here only
  represents the right/left part of the device. We need to glue them
  together to get the Nambu Green's function of the whole device
  \begin{equation}
  G_{xx}(i\omega_{n})=\left[i\omega_{n}-H_{x,x}^{isol}-\Sigma_{x-1,x-1}^{L}(i\omega_{n})-\Sigma_{x+1,x+1}^{R}(i\omega_{n})\right]^{-1}.
  \end{equation}
  Furthermore, we can also get
  \begin{align}
  G_{x+1,x}(i\omega_{n}) & =G_{x+1,x+1}^{R}(i\omega_{n})V_{x+1,x}G_{x,x}(i\omega_{n}),\\
  G_{x,x+1}(i\omega_{n}) & =G_{xx}(i\omega_{n})V_{x,x+1}G_{x+1,x+1}^{R}(i\omega_{n}).
  \end{align}
  And the Josephson current can be calculated as
  \begin{equation}
  I_{JJ}=\frac{ek_{B}T}{\hbar}\mathrm{Im}\sum_{\omega_{n}}\mathrm{Tr}\left[\tilde{V}_{x,x+1}G_{x+1,x}(i\omega_{n})-\tilde{V}_{x,x+1}^{\dagger}G_{x,x+1}(i\omega_{n})\right]
  \end{equation}
  with $\tilde{V}_{x,x+1}=\left(\begin{array}{cc}
  V_{x}^{ee}\\
   & -V_{x}^{hh}
  \end{array}\right)=\left(\begin{array}{cc}
  V_{x}^{ee}\\
   & (V_{x}^{ee})^{*}
  \end{array}\right)$. For a fixed external magnetic field, $I_{JJ}$ is a function of
  the pairing phase difference $\Delta\varphi$, thus we can get the flux-dependent current-phase relation $I_s(\Delta\varphi,\Phi)$.
  By varying $\Delta\varphi$ from $-\pi$ to $\pi$, we can also address the maximum magnitudes for both
  the positive and negative currents, which are the two critical Josephson currents along two opposite directions. In other words, the flux-dependent critical supercurrent $I_c(\Phi)$ presents the periodic Fraunhofer oscillations.

\begin{figure}[h]
  \centering
  \includegraphics[width=0.7\linewidth]{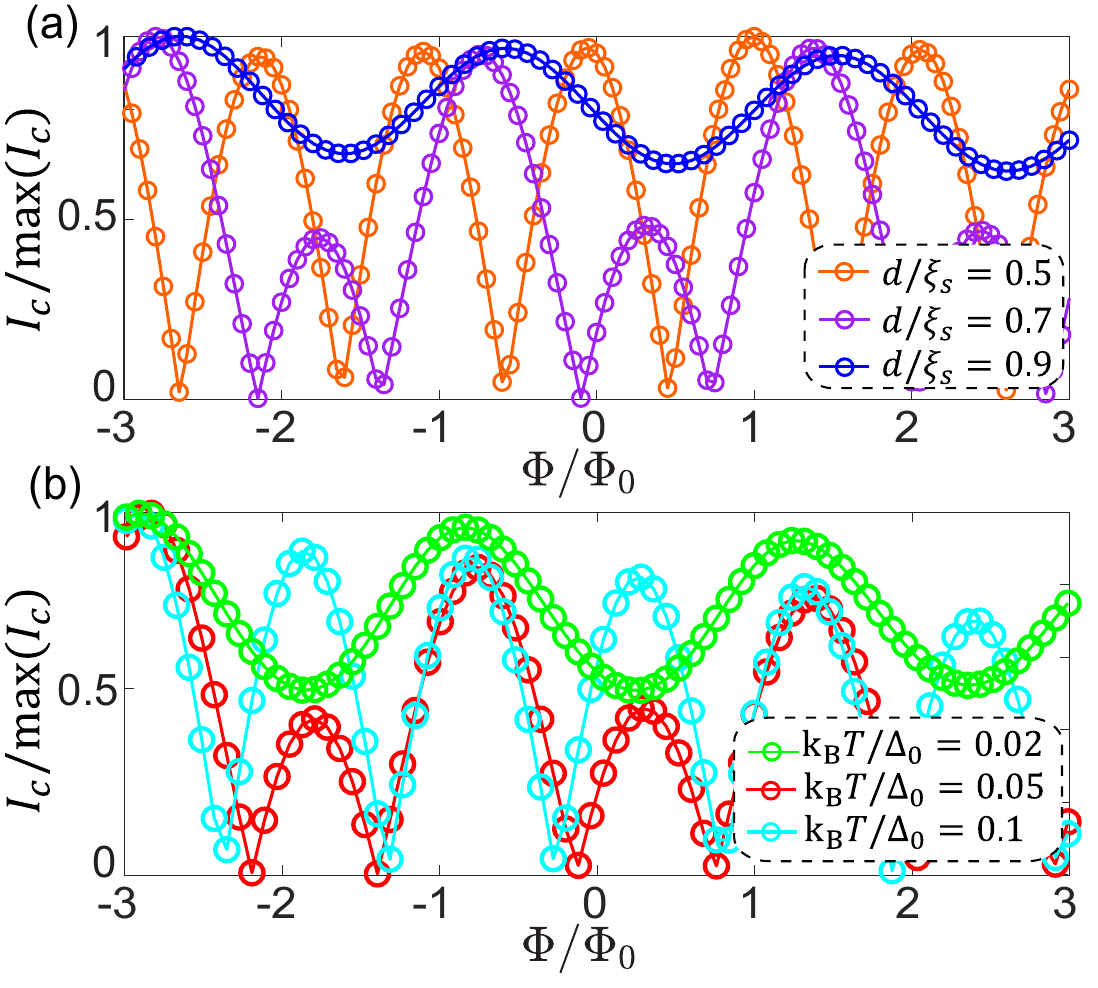}
  \caption{(a) and (b) show the crossover from $h/e$ oscillations to $h/2e$ oscillations by varying the junction length and temperature, respectively.}
  \label{fig:figS5}
\end{figure}

 \subsection{G. Some simulation results}

In this section, we simulate the Fraunhofer pattern numerically. Although similar simulations have been performed before \cite{li2022quantum}, the crossover between $h/e$ and $h/2e$ Fraunhofer oscillations has been overlooked. In Fig.~\ref{fig:figS5}~(a), numerically, we find the period of Fraunhofer oscillations gets transited from 2$\Phi_0$ to $\Phi_0$ as $d$ decreases, which is consistent with the analytical result in Fig.~\ref{fig:figS5}~(a). Numerical simulation with model parameters $(M,A,B)=(1,2,2)$. To be consistent with Fig.~\ref{fig:figS5}~(a), we fix the junction width $W=4\xi_s$, where $\xi_s=20a$ with $a=1$ is the lattice constant, and temperature $k_B T=0.05\Delta_0$ with $\Delta_0=0.1$, $\mu=0.1$, and $\mu'=1$, $t_s=1$, $t_c=0.7$.

In the actual experiments, changing the geometry of samples is not economical. Instead, to further experimentally verify our theory, we propose to detect a thermal-driven crossover between $h/e$ and $h/2e$ oscillations, which, as mentioned earlier, would be expected by our theory. In Fig.~\ref{fig:figS5}~(b), with fixed length $d=0.7\xi_s$ and width $W=4\xi_s$, we show that increasing the temperature will drive a $h/e$ to $h/2e$ Fraunhofer oscillations crossover.

For the simulation of Majorana zero modes, we directly diagonalize the lattice model with SC/CI/SC configuration, with parameters $(M,A,B)=(0.5,1,1)\text{eV}$, junction length $d=5a$, $a=1$ is the lattice constant, $\Delta_0=0.3\text{eV}$, $\mu=0.3\text{eV}$, and $\mu'=3\text{eV}$, $t_s=1\text{eV}$, $t_c=0.5\text{eV}$.

\end{document}